\documentclass[conference]{IEEEtran}
\newif\ifanon
\anonfalse
\ifdefined\buildanon \anontrue \fi
\newif\iflong
\longfalse
\ifdefined\buildlong \longtrue \fi

\usepackage{amsmath,amssymb}
\usepackage{graphicx}
\usepackage{booktabs}
\usepackage{array}
\usepackage{multirow}
\usepackage{xcolor}
\usepackage[colorlinks=true,linkcolor=blue,citecolor=blue,urlcolor=blue]{hyperref}
\usepackage{algorithm}
\usepackage{algpseudocode}
\usepackage{tikz}

\newcommand{\vpar}{v_\parallel}
\newcommand{\dd}{\mathrm{d}}

\begin{document}

\title{Parallel Design of a GPU-Resident Semi-Lagrangian Drift-Kinetic
Solver on Unstructured Extruded Tokamak Meshes}

\ifanon
\author{\IEEEauthorblockN{Anonymous submission}}
\else
\author{\IEEEauthorblockN{Mark F.~Adams}
\IEEEauthorblockA{Applied Mathematics and Computational Research Division\\
Lawrence Berkeley National Laboratory\\
Berkeley, CA, USA\\
mfadams@lbl.gov}
}
\fi

\maketitle

\begin{abstract}

Semi-Lagrangian (SL) schemes advect a distribution function along
characteristics and remap it to a fixed grid.  They are unconditionally
stable without a global solve, and with discontinuous Galerkin (DG)
integration the transfer along a magnetic field line conserves mass
exactly.  This paper presents the parallel and GPU design of a 3X+2V SLDG
drift-kinetic tokamak solver on unstructured, flux-surface-aligned
poloidal planes of shaped cross section, extruded toroidally.  Transport
follows magnetic field
lines across the planes, and each velocity channel advances a fixed
distance per step, proportional to its parallel speed, so the halo depth
a channel needs grows with its speed.
Velocity-dependent ragged halos exploit this, sending each channel only
its own upwind depth.  The 4D decomposition includes
the toroidal angle, the 2D poloidal plane and the magnetic moment, and
leaves the parallel velocity whole: no stage transports across the
magnetic moment, while a
parallel-velocity partition would put a halo inside the acceleration
stage.  On NVIDIA A100s the design strong-scales
over the magnetic-moment axis from one to 12 four-GPU nodes, and a weak series of three
points refining all five phase-space dimensions together, $16\times$ in
total per series point, holds a weak-scaling efficiency of $75\%$ on 16 GPUs and $57\%$
on 256, from a one-GPU base case.  The pointwise interpolations within the poloidal plane do not conserve
mass, and the full step's mass drift stays below
$2.1\times10^{-5}$ at every series point.  The design was implemented
with an AI code assistant under a machine-checked regression discipline,
in which every change is held to one of four exactness classes, from
bit-identical to a physics rate within a stated tolerance.
\end{abstract}


\section{Introduction}
\label{sec:intro}

Kinetic plasma simulation resolves the distribution function
$f(\mathbf{x},\mathbf{v},t)$ on a phase-space grid, and its cost is
dominated by the transport of $f$ and by the velocity-space
dimensionality.  Semi-Lagrangian (SL) schemes integrate transport along
characteristics and are stable for time steps far beyond the CFL limit
of explicit discretizations, and combined with a discontinuous Galerkin
(DG) representation a shift along a grid axis conserves mass exactly
through $L^2$ projection onto the shifted solution~\cite{einkemmer,crouseilles}.  The
large-$\Delta t$ advantage restructures the parallel communication
problem.  In a tokamak the transport is along magnetic field lines,
which cross the poloidal planes of an extruded mesh, so a channel of
parallel speed $\vpar$ reads $\vpar\,\Delta t$ worth of upstream planes
per step and the halo a rank requires grows with $\Delta t$ and with the
fastest resolved parallel speed.  A fixed-width halo sized for
$v_{\max}$, the textbook pattern and what a distributed-array library
provides by default, moves more data than the stencil reads, because
most channels move slower than $v_{\max}$ and each needs a halo only on
its upwind side.  The halo must also fit in one neighbour's block, so
the toroidal angle alone offers at most one rank per halo width around
the torus (Sec.~\ref{sec:ceiling}).

After the toroidal axis the next design choice is which axis to cut, or
not cut.  Configuration space is
cut in any distributed decomposition and pays the halos of the transport
across it; the open choice is in velocity space.  No stage of the
operator transports across the magnetic moment, so that axis costs no
exchange to partition and is cut.  The acceleration stage, the velocity
kick, transports along the parallel velocity, so cutting that axis would
put a halo inside the kick, and a velocity grid is small enough that the
parallel velocity can be left whole.

This paper describes the parallel and GPU design of KINTORA, a
continuum SLDG drift-kinetic solver on unstructured triangular poloidal
planes extruded toroidally, built on PETSc~\cite{petsc} and
Kokkos~\cite{kokkos}.  The closest prior work, the SLDG code line of
Einkemmer et al.~\cite{einkemmer2016,einkemmer2020,einkemmer2023},
scales uniform-grid SLDG to large GPU machines by decomposing all
phase-space dimensions, and it has run
drift-kinetic simulations of the ion-temperature-gradient (ITG)
instability on a structured cylindrical mesh on a single
GPU~\cite{einkemmer2022}.  The present code targets two things:
shaped, unstructured poloidal planes and distributed GPU machines.  The
method, its physics
and its verification against independently computed references are
discussed in more detail in a companion paper~\cite{companion}; the
present paper is the code's parallel design and the measurements of that design.  This paper's
contributions are:

\begin{itemize}
\item \textbf{A four-dimensional decomposition that leaves the parallel
velocity whole} (Secs.~\ref{sec:ragged}--\ref{sec:field}): it divides
the toroidal angle, the poloidal plane and the magnetic-moment ($\mu$)
rows of a step whose state never leaves the GPU.  The $\mu$ rows, which
no stage couples, multiply the toroidal rank count, which the deep halo
caps, by the number of rows (Sec.~\ref{sec:rows}).  The toroidal halo exchange is device-direct,
overlapped with interior work and sent at each channel's own upwind
depth, and the per-plane field solves are divided among the row ranks.
\item \textbf{Support for unstructured, shaped poloidal planes}
(Secs.~\ref{sec:solver} and~\ref{sec:field}): the parallel stream traces each
field line of a shaped equilibrium and evaluates $f$ at the line's foot
in an unstructured triangle, the plane decomposition cuts the mesh into bands of
poloidal flux, and each plane's polarization equation is factored once
and solved directly, with a flux-surface-averaged electron response.
\item \textbf{Measurements on Perlmutter A100 GPUs}
(Sec.~\ref{sec:eval}): strong scaling over the $\mu$ row axis to 48 GPUs, and
an accuracy-matched weak series on the shaped plane that refines all
five phase-space dimensions together, $16\times$ in total per series point,
to 256 GPUs, with what each named optimization is worth and a roofline
profile of the step's dominant kernels.
\item \textbf{A machine-checked regression discipline for AI-assisted
implementation} (Sec.~\ref{sec:method}): every change is held to one of
four exactness classes, from bit-identical to a physics rate within a
stated tolerance, asserted against committed reference values on the CPU
and GPU builds.
\end{itemize}

Conservation is the known weakness of the semi-Lagrangian class, and
this code is not an exception.  The transfer along each field line and
the shift along $\vpar$ conserve mass exactly under the $L^2$ projection.
A dilation in $\vpar$, a rescaling about $\vpar = 0$ that carries the
curvature drift's energy exchange with the field, changes the mass by an
amount that, in the continuum, cancels the compressibility of the
E$\times$B flow.  The stream's evaluation where a field line
crosses a station ($\times$ in Fig.~\ref{fig:torus}) and the evaluation
of the two in-plane stages, the drift and the E$\times$B, at a departure
foot within the plane are pointwise interpolations on unstructured
triangles, and do not conserve mass.  A conservative in-plane remap is
in progress; the companion paper reports one for the drift
stage~\cite{companion}.

Sec.~\ref{sec:solver} states what the parallel design depends on: the
mesh, the velocity grid, the Strang step and the stream's stencil.
Secs.~\ref{sec:ragged} to~\ref{sec:field} are the design, and
Sec.~\ref{sec:method} the machine-checked exactness gates used for
regression testing.  Sec.~\ref{sec:eval} evaluates the design on
Perlmutter A100 nodes: strong scaling over the $\mu$ row axis to 48 GPUs,
the accuracy-matched weak series, what each named optimization is worth,
and where each stage's dominant kernel sits on the roofline.

\section{Background: the tokamak solver}
\label{sec:solver}

\begin{figure}[t]
\centering
\resizebox{\columnwidth}{!}{\input{fig_torus_pencil_body.tex}}
\caption{The advection stream of one channel $j$.  Everything in
red is the backward trace: node $q$'s field line is traced back (solid)
to its foot, drawn at a shift of $1.6$ wedges ($n = 1$ and $\alpha = 0.6$
in Eq.~\eqref{eq:shift}), and the arrow is the distance the trace covers in
a half step, $v_{\parallel,j}\, g_q \Delta t/2$ along the line, with
$g_q = \dd\varphi/\dd\ell$ the line's rate factor.  The same trace takes wedge $c$ (dashed)
to its pre-image (shaded), overlapping wedges $c{-}1$ and $c{-}2$, and
the transfer integrates both overlaps.  Where the line crosses a station
it lands in a triangle ($\times$) with six $P_2$ weights.  The solid
trace from $q$ to its foot is the \emph{pencil} of channel $j$ at node
$q$.}
\label{fig:torus}
\end{figure}

The solver advances the drift-kinetic equation for ions with adiabatic
electrons and a long-wavelength polarization field on a phase space of
two parts: an unstructured triangular poloidal plane crossed
with the toroidal angle $\varphi$ (Fig.~\ref{fig:torus}), and the
gyrotropic velocity space $(\vpar, \mu)$ of parallel speed and magnetic
moment (Fig.~\ref{fig:ragged}).  The plane of Fig.~\ref{fig:torus} is
the geometry the solver targets, a flux-surface-aligned mesh
of an up-down asymmetric analytic Solov'ev
equilibrium~\cite{cerfon2010} (inverse aspect ratio $0.36$, elongation
$1.7$ and $1.9$, triangularity $0.25$ and $0.45$, no X-point), on which
every vertex lies on a flux surface.  The weak series of
Sec.~\ref{sec:weak} is timed on this plane at three refinements (levels
4, 5 and 6) and the
strong series over the $\mu$ row axis (Sec.~\ref{sec:rowscale}) at the first
of them.

\begin{figure}[t]
\centering
\resizebox{\columnwidth}{!}{\begin{tikzpicture}[font=\footnotesize, >=stealth,
  x={(0.30cm,0cm)}, y={(0cm,0.22cm)}, z={(0.21cm,0.21cm)},
  d1/.style={fill=red!10}, d2/.style={fill=red!30}, d3/.style={fill=red!52},
  own/.style={draw=black!55, line width=0.3pt, fill=blue!8},
  halo/.style={draw=red!60!black, line width=0.3pt},
  env/.style={draw=black!60, line width=0.5pt, dashed},
  lab/.style={font=\tiny, inner sep=1pt},
  tag/.style={font=\tiny, inner sep=1pt, text=black!70}]
  \def\chans{0/3/1/1, 1.505/3/1/1, 3.939/2/1/1, 5.444/1/1/2, 6.150/1/1/1, 7.293/1/1/1, 8.000/1/0/2,
             8.707/1/0/1, 9.850/1/0/1, 10.556/1/0/2, 12.061/2/0/1, 14.495/3/0/1, 16.000/3/0/1}
  \def\hb{0.22} %
  \def\nodex{0, 1.505, 3.939, 5.444, 6.150, 7.293, 8.000, 8.707, 9.850, 10.556, 12.061, 14.495, 16.000}
  \def\rowy{0.000, 0.146, 0.292, 0.520, 0.748, 1.104, 1.459, 2.014, 2.569, 3.434, 4.300, 5.650, 7.000}

  \foreach \x/\d/\sd/\nc in \chans {
    \ifnum\nc=2 \def\lays{-0.32,0}\else \def\lays{0}\fi
    \foreach \yy in \lays {
      \foreach \k in {-2,...,1}
        \draw[own] (\x-\hb,\yy,\k) -- (\x+\hb,\yy,\k) -- (\x+\hb,\yy,\k+1) -- (\x-\hb,\yy,\k+1) -- cycle;
      \foreach \k in {1,...,\d} {
        \ifnum\sd=1
          \draw[d\d, halo] (\x-\hb,\yy,\k+1) -- (\x+\hb,\yy,\k+1) -- (\x+\hb,\yy,\k+2) -- (\x-\hb,\yy,\k+2) -- cycle;
        \else
          \draw[d\d, halo] (\x-\hb,\yy,-\k-2) -- (\x+\hb,\yy,-\k-2) -- (\x+\hb,\yy,-\k-1) -- (\x-\hb,\yy,-\k-1) -- cycle;
        \fi
      }
    }
  }
  \draw[env] (-\hb,0,2) -- (7.293+\hb,0,2) -- (7.293+\hb,0,5) -- (-\hb,0,5) -- cycle;
  \draw[env] (8-\hb,0,-5) -- (16+\hb,0,-5) -- (16+\hb,0,-2) -- (8-\hb,0,-2) -- cycle;
  \draw[black!60, line width=0.5pt] (16.3,0,-2) -- (16.55,0,-2) -- (16.55,0,2) -- (16.3,0,2);
  \node[tag, anchor=east] (ncl) at (5.90cm,-0.084cm) {$n_{cl} = 4$};
  \node[tag, anchor=north east] at (ncl.south east) {owned wedges};
  \draw[->, black!70, line width=0.4pt] (16.8,0,-5.6) -- (16.8,0,4.1) node[anchor=south east, tag] {$\varphi$};

  \foreach \y in {0.000, 0.292, 0.748, 1.459, 2.569, 4.300, 7.000} \draw[black!45, line width=0.3pt] (0,\y,0) -- (16,\y,0); %
  \foreach \x in {0,5.444,8,10.556,16} \draw[black!45, line width=0.3pt] (\x,0,0) -- (\x,7,0);
  \foreach \x in \nodex \foreach \y in \rowy \fill[black!80] (\x,\y,0) circle (0.016cm);
  \node[tag, anchor=east, align=center] at (-0.3,3.5,0) {$\mu$ rows\\($v_\perp$)};

  \draw[->, black!70, line width=0.4pt] (0,8.4,0) -- (16,8.4,0) node[right, tag] {$\vpar$};
  \node[tag, anchor=south] at (8,8.6,0) {4 segments of degree 3, graded to $\vpar = 0$, GLL nodes};
  \node[tag, anchor=north east] at (7.6,-5.4,0) {$\vpar < 0$: halo on $+\varphi$ (behind)};
  \node[tag, anchor=north west] at (8.4,-5.4,0) {$\vpar \ge 0$: halo on $-\varphi$ (in front)};
  \coordinate (stk) at (8-\hb,-0.16,-2.75);
  \draw[->, black!70, line width=0.4pt] ([xshift=-0.55cm]stk) -- ([xshift=-0.08cm]stk);
  \node[tag, anchor=east, align=right] at ([xshift=-0.57cm]stk) {stack where two\\ GLL nodes meet};
  \begin{scope}[shift={(-1.0,-8.2,0)}]
    \draw[d2, halo] (0,0,0) rectangle (1.2,1,0);
    \node[lab, anchor=west] at (1.4,0.5,0) {ragged: the channel's own depth, on its upwind side};
    \draw[own] (0,-1.6,0) rectangle (1.2,-0.6,0);
    \node[lab, anchor=west] at (1.4,-1.1,0) {owned};
    \draw[env] (0,-3.2,0) rectangle (1.2,-2.2,0);
    \node[lab, anchor=west] at (1.4,-2.7,0) {one-sided worst case: every channel at $s_w = 3$};
  \end{scope}
\end{tikzpicture}}
\caption{The ragged halo of one of four toroidal ranks, which owns four
of the 16 wedges, at a step whose fastest channel shifts $s_w = 3$
wedges; the setup admits $s_w \le N_\varphi/P_\varphi$.  Velocity space
($\vpar$ across, $\mu$ up) stands as a wall on the toroidal angle
$\varphi$, which runs into the page; $\vpar$ is not partitioned.  Under each $\vpar$ node a strip
depicts the one-dimensional grid of wedges along $\varphi$: the rank's
four owned wedges,
centred on the wall, and the upwind halo, behind the wall for
$\vpar < 0$ and in front of it for $\vpar \ge 0$.  The halo's depth is set
by $\vpar$ and the field line's rate factor $g_q$ of
Eq.~\eqref{eq:shift} and is the same in every $\mu$ row.}
\label{fig:ragged}
\end{figure}

\subsection{Discretization}
\label{sec:disc}

The poloidal plane carries nodal Lagrange elements of degree $p$ ($P_2$
in production) on $N_{\mathrm{pol}}$ nodes.  The toroidal direction is
$N_\varphi$ DG elements, the \emph{wedges}, each carrying the
$p_\varphi{+}1 = 3$ Gauss--Lobatto--Legendre (GLL) stations of degree
$p_\varphi = 2$.  A station is a copy of the plane.  Velocity space
is a tensor grid, DG in $\vpar$ ($N_v$ \emph{segments} of degree $p_v$,
one-dimensional elements with GLL nodes) by $N_{\mathrm{row}}$
\emph{$\mu$ rows}, lines of constant $\mu$.  The rows are quadrature
points, not elements, since no operator couples them: the GLL nodes of a
one-dimensional grid on the coordinate $w = \sqrt{2\mu B_0}$, which is
$\mu$ expressed as the perpendicular speed at the reference field $B_0$
($v_\perp = w\sqrt{B/B_0}$ elsewhere).  Segments uniform in $w$ give
rows graded in $\mu$, finest at $\mu = 0$.  The word cell is used for the
triangles of the plane; a triangle crossed with a wedge and a $\vpar$
segment, at one $\mu$ row, is the phase-space cell of the method.  A
\emph{channel} is one $\vpar$ node on one $\mu$ row, a single velocity point,
whose data span every spatial node, and
$N_{\mathrm{ch}} = N_v(p_v{+}1)N_{\mathrm{row}}$ is the number of
channels, the size of the velocity grid; Fig.~\ref{fig:parallel} draws a
velocity mesh of this kind and Table~\ref{tab:weak} gives the grid at
each series point.  The state holds one double for each channel at
each node of each station, $3N_\varphi N_{\mathrm{pol}} N_{\mathrm{ch}}$
doubles in all.

We grade the tensor-product velocity grid to gain accuracy at a given
number of points.  Further grading of the $\mu$ rows did not reduce the
error.  In $\vpar$ we find that segments that each hold
an equal share of the mass of a Maxwellian 2.5 to 3 times the thermal
width reduce the error in the growth rate of a linear ITG mode from
$7\%$ to under $1.7\%$, and in its frequency from $4\%$ to $1.1\%$,
against uniform segments at 32 $\vpar$ nodes by 32 $\mu$ rows.  Each
error is the difference from a $64\times48$ grid; a $48\times48$ grid
differs from that reference by up to $3.4\%$ in the rate, so the
comparison does not resolve differences much below that.  We measured these errors on the circular plane at the same
refinement with 16 wedges, at a steeper temperature gradient than the
runs timed in Sec.~\ref{sec:eval}, where the mode grows cleanly; on the shaped plane that mode is not yet
characterized.

The stepper is Strang splitting:
the parallel stream for $\Delta t/2$; the grad-$B$ and curvature drift
along each node's vertical fiber for $\Delta t/2$; the zeroth velocity moment of $f$
at every station, the density, corrected by a midpoint density
predictor; the polarization solve on each plane, with the
flux-surface-averaged electron response (Sec.~\ref{sec:field}); the
velocity kick, one remap along $\vpar$ that composes $E_\parallel$ and
the mirror force into a shift and the curvature drift's energy exchange
with the field into a dilation; the E$\times$B term as a semi-Lagrangian stage
in the plane, which evaluates $f$ at an in-plane foot of each node; the
second half-drift; and the second half-stream.  This is the stepper every measurement of
this paper runs (Sec.~\ref{sec:eval}).  Each stage moves content along
one direction of phase space: the two half-streams along the field line,
the kick along $\vpar$, the E$\times$B stage and the drift within the
plane; see the companion paper for details~\cite{companion}.  None moves content across $\mu$, which is invariant along a
field line and under the drift, so every stage is a per-$\mu$-row
operation, a structure of the equations that the $\mu$ row axis of
Sec.~\ref{sec:rows} exploits.  Collision operators, which the present
model omits, mix the $\mu$ rows and would require a transpose of the state
across the $\mu$ row axis, a substantial cost but subdominant to that of
an accurate collision operator~\cite{adamslandau2025,taitano2015}.

\begin{figure*}[t]
\centering
\resizebox{\textwidth}{!}{\input{fig_parallel48_body.tex}}
\caption{The decomposition on 24 ranks: $4$ toroidal blocks $\times$ $2$
pieces of the poloidal plane $\times$ $3$ blocks of $\mu$ rows.  A
\emph{wedge} is the DG element in $\varphi$ and carries three
\emph{stations}, each a full copy of the plane.  The key numbers the five
recurring communications and names the ranks each one spans.  The velocity mesh
at the right of each row of ranks is the mesh at any one spatial point,
and its shaded box is that row's $\mu$ block.  All three planes are the shaped cross
section the timed runs use, (c) cut as the partition of
Sec.~\ref{sec:parallel} cuts it.  Insets (a) and (b) are drawn with the
plane in $P = 4$ pieces, where a stream halo exists; at $P = 2$ both pieces are annuli.}
\label{fig:parallel}
\end{figure*}

\subsection{The stream}

The stream is the communication-bearing stage, and it is the
flux-coordinate independent construction of Hariri and
Ottaviani~\cite{hariri2013} and of GRILLIX~\cite{stegmeir2018}: the
field line is traced exactly and the plane is interpolated at its foot,
so the mesh is never aligned with the field.  For channel $j$ on the
field line through node $q$, the shift along the line is
\begin{equation}
\label{eq:shift}
s_{j,q} \;=\; \frac{v_{\parallel,j}\, g_q\, \Delta t}{2\,\Delta\varphi}
\;=\; n + \alpha
\end{equation}
wedges, with $g_q = \dd\varphi/\dd\ell$ the line's rate factor, integer
$n$ and fractional $\alpha\in[0,1)$.  The update in wedge $c$ is the
two-wedge stencil
\begin{equation}
\label{eq:xupdate}
f^{+}_{c} \;=\; A(\alpha)\, f_{c-n} \;+\; B(\alpha)\, f_{c-n-1},
\end{equation}
where $A$ and $B$ are the $3\times3$ overlap matrices of the GLL basis
at shift $\alpha$~\cite{einkemmer}, and the field line lands inside a
triangle of the upstream plane, so the in-plane part of the map is a
pointwise evaluation of the nodal basis at that point ($\times$ in
Fig.~\ref{fig:torus}).
Equation~\eqref{eq:shift} is the closed form the circular concentric
equilibrium admits, where $g_q$ is constant along a line; on the shaped
equilibrium the stream traces the line and root-finds the shift along
it.  Neither form depends on time, so $n$, $\alpha$ and the plane maps
are static and are built once.  The stencil reads wedges $c-n-1$
and $c-n$, on the upwind side only.  A channel restricted to one poloidal
node, the line of points along $\varphi$ that the stencil shifts, is a
\emph{pencil}, and a channel's halo depth is the largest $n+1$ over its
pencils, because $g_q$ varies from line to line.  Mass conservation along the
line is algebraically exact: the GLL basis is a partition of unity and
$A$ and $B$ are its overlap integrals, so the weighted column sums of
$A+B$ return the source weights.

\subsection{Parallel structure}
\label{sec:parallel}

The decomposition is four-dimensional (Fig.~\ref{fig:parallel}).  The
toroidal axis partitions the wedges: a rank owns
$n_{cl} = N_\varphi/P_\varphi$ consecutive wedges and replicates the
poloidal and velocity metadata.  The plane axes partition the poloidal
mesh into $P$ pieces: bands of poloidal flux and, where a band is long against its
thickness, into sectors within that band.  These pieces divide the state
and the in-plane stencils, while the field solve keeps each plane whole
(Sec.~\ref{sec:field}); a plane cut into whole annuli (Fig.~\ref{fig:parallel}(c)) uses the
radial direction alone, and the decomposition is then three-dimensional.  The row axis partitions the
$\mu$ rows over $P_{\mathrm{row}}$ blocks (Sec.~\ref{sec:rows}).  The
parallel velocity is the one phase-space direction no axis divides, and
Sec.~\ref{sec:ragged} gives the reason.  All state is device resident
(Kokkos~\cite{kokkos}, PETSc~\cite{petsc}), and the halo exchanges and the
collectives over the $\mu$ rows pass device buffers to MPI
(Sec.~\ref{sec:ragged}).  Five
communications recur: (i) the stream's halo exchange between toroidal
neighbours; (ii) a depth-one scalar halo of the potential for the
parallel gradient; (iii) one all-reduce of the density's channel sum
over the $\mu$ row ranks; (iv) one all-gather of the field over the row ranks;
and (v) one all-reduce of $K$ surface sums for the flux-surface average,
$K$ the number of flux surfaces of the mesh (17, 33 and 65 at the three
series points).  Secs.~\ref{sec:ragged}
to~\ref{sec:field} describe them in turn.

Two properties frame the design.  First, the halo is \emph{deep relative
to the block}.  The halo width $s_w$ is the per-step displacement of the
fastest channel in wedges, and one wedge is three copies of the plane
times every channel, so at four wedges per rank, the shape of every
series point, a halo of one wedge is a quarter of the owned
volume on each side.  Second, a fixed-width exchange \emph{oversizes}
the halo, because a channel reads $|n|+1$ wedges on one side while the
fastest channel sets $s_w$.  At every series point a channel's depth
and $s_w$ coincide by design: the series halves $\Delta t$ as it doubles $N_\varphi$, so
by~\eqref{eq:shift} the largest shift is the same fraction of a wedge at
every series point and under one wedge, so every channel has $n = 0$ and depth
one (Fig.~\ref{fig:torus} draws $n = 1$, depth two), and the per-channel
depths equal the one-sided worst case.  Once the shift passes a wedge, at
a larger $\Delta t$ or at more wedges for the same $\Delta t$, the gains
of ragged halos (Sec.~\ref{sec:ragged}) are realized.

\section{Velocity-dependent ragged halos}
\label{sec:ragged}

\subsection{Upwind-only per-channel depths}

From~\eqref{eq:xupdate}, channel $j$ reads wedges $c-n-1$ and $c-n$ on
every line, so with $n_j$ the largest shift over the lines of the
plane it needs
\begin{equation}
\label{eq:depth}
d_L(j) = \begin{cases} n_j+1, & n_j\ge 0\\ 0,& n_j<0\end{cases}
\qquad
d_R(j) = \begin{cases} 0, & n_j\ge 0\\ -\,n_j, & n_j<0\end{cases}
\end{equation}
halo wedges on its $-\varphi$ side ($d_L$) and its $+\varphi$ side
($d_R$), nonzero on one side only, determined by the sign of its
parallel speed.  The baseline is a fixed-width exchange that sends every channel
the worst-case depth $s_w$ on its upwind side, for a volume of
$s_w N_{\mathrm{ch}} N_{\mathrm{blk}}$, with
$N_{\mathrm{blk}} = 3N_{\mathrm{pol}}$ the doubles one channel holds per
wedge.

The per-channel depths of~\eqref{eq:depth} give a volume
$\sum_{j}\big(d_L(j)+d_R(j)\big)N_{\mathrm{blk}}$ against
$s_w N_{\mathrm{ch}} N_{\mathrm{blk}}$, so the gain is the ratio of the
worst-case depth to the mean depth.  The gain is a property of the shift
distribution, which $\Delta t$ and $N_\varphi$ set.  It is $1.00$
wherever every channel shifts by under a wedge, as at every series point
(Sec.~\ref{sec:solver}), and as the time step grows it approaches the
ratio of $v_{\max}$ to the mean $|\vpar|$: $2$ on a uniform $\vpar$ grid,
and more on a grid graded or adapted toward $\vpar = 0$, which holds more
slow channels.  The most accurate of the $\vpar$ grids compared in
Sec.~\ref{sec:disc}, equal-mass segments, is of this kind.  Fig.~\ref{fig:ragged} draws this grid with four
segments, at three times the thermal width.  The
one-sided worst case is also available as a run mode, bit-identical to
the ragged one in its physics, and Sec.~\ref{sec:raggedtime} measures
what the two cost in time at a shift past one wedge.

Sec.~\ref{sec:intro}'s reason against partitioning $\vpar$ is the kick,
which shifts content along $\vpar$ by an
amount proportional to $E_\parallel - \mu\,\partial_s B$, so a $\vpar$
partition would put a halo inside it.  A $\mu$ block holds the full
range of parallel speeds, so every rank has the same depths, and the
ragged exchange reduces every rank's volume by the same ratio.

\subsection{Ragged layout, device staging, persistent requests}

The variable depths make the send and receive buffers ragged.  Element
$(j,k,b)$, channel $j$, halo wedge $k$ and block entry $b$, lives at
$\mathrm{off}[j] + kN_{\mathrm{blk}} + b$ with host-side prefix sums
$\mathrm{off}[j]=\sum_{j'<j} d(j')N_{\mathrm{blk}}$ mirrored to the
device.  A pack kernel gathers the upwind wedges of every channel into
device staging buffers of these sizes, and the MPI requests are bound to
the staging buffers themselves, which GPU-aware MPI reads and writes in
device memory (Cray MPICH with \texttt{MPICH\_GPU\_SUPPORT\_ENABLED=1});
an unpack kernel scatters the received buffers into the ghost wedges each
channel reads.  Because depths, hence counts, are fixed for a run, all
four transfers, a send and a receive to each neighbour, use persistent MPI requests
(\texttt{MPI\_Send\_init}/\texttt{Recv\_init} once,
\texttt{MPI\_Startall}/\texttt{Waitall} per exchange), batching both
directions non-blocking.  Staging the buffers through pageable host
memory instead, one bulk copy of each buffer, gives bit-identical results and a
step $2.35\times$ longer at point C (Table~\ref{tab:ablation}).  The
depth-one scalar halo of the potential is a second instance of the same
exchange, with one channel.\iflong{}
Sec.~\ref{sec:onex} reports the class's cost in the 1X+3V solver, where
it was first measured.\fi

\subsection{Overlapping the exchange with interior work}
\label{sec:overlap}

Wedges $c\in[\max_{j} d_L,\; n_{cl}-\max_{j} d_R)$ read no halo data
for any channel.  The stream therefore launches in three pieces.  The
first is the pack and \texttt{MPI\_Startall}.  The second is the
interior stream, launched asynchronously so the GPU computes while the
host waits in \texttt{MPI\_Waitall}.  The third is the unpack and the
two boundary strips.  Each output value is computed by exactly one
thread with unchanged arithmetic, so the split is bit-identical to the
monolithic kernel.

The in-plane exchanges between the pieces of a cut plane overlap the
same way.  The drift, the E$\times$B stage and the density moments order
their targets interior first, interior meaning that every node a
target's stencil reads is owned, and launch in two pieces: the pack and
the start precede the interior launch, and the wait and the unpack
precede the boundary one.  The drift's order is fixed at setup; the
E$\times$B stage's feet move with $v_E$, so its order is rebuilt by a
prefix scan in every foot pass.  The stream overlaps only the owned part
of its gather, because the toroidal message carries the plane's ghost
nodes and cannot be packed until the in-plane exchange has arrived.  These
splits are bit-identical as well.

\subsection{The toroidal rank ceiling}
\label{sec:ceiling}

The exchange is nearest-neighbour, so the halo width must fit in one
neighbour's block, $s_w \le n_{cl} = N_\varphi/P_\varphi$.  The halo
width is the per-step displacement of the fastest channel in wedges,
and at fixed $\Delta t$ it grows with $N_\varphi$: by~\eqref{eq:shift}
a configuration that needs $s_w = 1$ at $N_\varphi = 16$ needs
$s_w = 2$ at $N_\varphi = 32$, where 32 ranks would leave one wedge per
rank against a halo width of two, so the exchange would read past its
neighbour into the block beyond it; the setup checks $s_w \le n_{cl}$
and stops a run that violates the bound.  The usable toroidal rank count is $N_\varphi/s_w$,
and at fixed $\Delta t$ it rises only in steps, when the fastest foot
passes another wedge: in the configuration of Sec.~\ref{sec:raggedtime},
16 wedges admit 16 toroidal ranks, 32 wedges still 16, and 64 wedges 21,
so raising $N_\varphi$ to spread a run wider gains little.
Halving $\Delta t$ as $N_\varphi$ doubles, as the weak series does,
holds $s_w$ at one wedge and the ceiling at $N_\varphi$ toroidal ranks; the
series uses $N_\varphi/4$ of them, four wedges per rank.  These toroidal
counts fall far short of the series' rank counts: its third point runs on 256 GPUs with
16 wedges and its fourth on $4{,}096$ with 32.  Memory sets the same
limit from the other side: the fourth point's state is
$4.3\times10^{11}$ doubles, $3.4$\,TB, or $107$\,GB per rank over 32
toroidal ranks, more than one A100 holds.  Both limits are removed with
the $\mu$ and plane decomposition.

\section{The velocity-row axis}
\label{sec:rows}

No stage of the step couples the $\mu$ rows (Sec.~\ref{sec:solver}), so
they are the next axis to distribute.  The communicator splits
into a toroidal and a row communicator: the $\mu$ rows are partitioned
over the row ranks and the toroidal decomposition is unchanged over the
toroidal ranks.  The
velocity grid is built globally and sliced, so the per-channel tables
of every stage address a rank's own $\mu$ rows with no change.  The one
quantity a rank can no longer form alone is the density, the sum over
all channels, so the axis adds one all-reduce of
$n_{sp} = 3n_{cl}N_{\mathrm{pol}}$ doubles per step, the field's size and
$1/N_{\mathrm{ch}}$ of the state: at the second point of the weak
series, $n_{sp} = 12\times33{,}281$ doubles, or $3.2$\,MB, against a
$6.1$\,GB state per rank at one row block.  The axis multiplies the
usable rank count by the number of row blocks, 8 at this second point
and 32 at the third, and divides the per-rank state by the same
factor.  Fig.~\ref{fig:parallel} draws all four axes and every recurring
communication with the ranks it spans and its size.  The strong-scaling series
of Sec.~\ref{sec:eval} is timed over the toroidal and row axes, and the
weak series adds the poloidal plane decomposition ($P = 2$) at its
third point, 256 GPUs as $4 \times 2 \times 32$ (Table~\ref{tab:weak}).

Two properties of the implementation are asserted by the gate of
Sec.~\ref{sec:method}, the discipline every parallel change in this
project is held to.  One row block is byte-identical to the single-axis
path.  Across $P_\varphi = 1, 2, 4$ and $P_{\mathrm{row}} = 1, 2, 4$ the
peak field, mass, kinetic energy and field energy histories agree with
the one-block run to a worst relative deviation of $2.7\times10^{-12}$:
machine precision rather than bit-identical, and the mechanism is
specific.  The row all-reduce changes the order in which the density's
channel sum accumulates, and $\delta n = (n - n_0)/n_0$ is a
cancellation, so a relative move of $10^{-16}$ in $n$ is one of
$10^{-13}$ in $\delta n$ before the field solve amplifies it.  Every new
axis is held to the same pair: bit-identical at the axis's size one,
machine precision as it scales.

The $\mu$ rows are partitioned by count, not by work, and one stage's
work is not uniform across rows: the drift displacement is proportional
to $\vpar^2 + \mu B$, so a high-$\mu$ row's feet lie farther from their
nodes and its halo is wider.  That non-uniformity is subdominant here: the drift
stage is $10\%$ of the step at one row block and $7\%$ at twelve
(Sec.~\ref{sec:eval}), and the profile's ratio of the slowest rank to
the fastest is $1.0$ for every event at every row count, so a weighted
partition has nothing to redistribute.  A practical ramification of the
$\mu$ decomposition is that the block count must divide the number of
$\mu$ rows exactly.  What bounds the axis in practice is measured in Sec.~\ref{sec:eval},
and it is the field path, not the reduction.

\section{The field path}
\label{sec:field}

\subsection{Per-plane solves, divided among the row ranks}

The polarization equation is one two-dimensional system per station,
and the stations are independent.  Each system is solved by a sparse
direct Cholesky solver (NVIDIA cuDSS), factored once, on a single rank,
so no solve communicates.  The row ranks of a toroidal rank divide its
$3n_{cl}$ stations among them in contiguous slices, and one all-gather of $n_{sp}$
doubles over the row ranks, the payload of the density all-reduce, then
gives every rank the whole field.  A rank whose slice holds several
stations stacks them into one block-diagonal system and solves it in one
call: at point A one rank solves all 12 stations this way, and at point B
each of 8 row ranks solves one or two.  Where the row ranks outnumber the
stations, a rank solves one plane or none: at point C, where a toroidal
rank's four wedges hold 12 stations, 12 of the 32 row ranks solve and
the other 20 only receive the field.  At every series
point the solve is about $1\%$ of the step (Table~\ref{tab:weak}).
With one plane per rank the solve takes a single plane's time,
$0.53$\,ms at twelve blocks (Table~\ref{tab:rows}), which more row
ranks cannot lower.  Every solve runs on one GPU, because no plane
we run is large enough for a distributed solve to pay: the direct solve
is latency-bound, and a plane four times the size of point C's solves in
$1.2\times$ the time of point C's.

\subsection{The flux-surface-averaged electron response}
\label{sec:fsa}

The field equation is the long-wavelength gyrokinetic quasi-neutrality
condition~\cite{lee1983,hahm1988}, whose adiabatic electron response
subtracts the flux-surface average of $\phi$~\cite{hammett1993}, which
couples every station through $K$ surface sums.  With the
$K$ averages as unknowns beside $\phi$, the field equation is a
$2\times2$ block KKT system, solved by a direct block $LDL^T$
factorization with an explicit Schur complement: a dense $K\times K$
matrix formed at setup from $K$ solves with the per-plane operator and
factored once.  Each step then adds a correction to the per-plane solves:
one all-reduce of $K$ doubles over the ranks that hold one row block, a
dense $K\times K$ solve on every rank for the $K$ surface averages of
$\phi$, and a field built from those averages added to every station.  Timed as its own event on A100, the
correction costs $0.17$, $0.35$ and $1.11$\,ms per step at the three
series points ($K = 17$, 33 and 65), $6.6\%$, $7.6\%$ and $11.7\%$ of the
field stage of Table~\ref{tab:weak}.  The growth with $K$ is likely the
dense $K\times K$ solve: the surfaces roughly double from point to point
and the cost rises $2.1\times$ and then $3.1\times$.  The plane and the
rank shape change with each point as well, so the growth is not $K$
alone.

\section{Machine-checked regression discipline}
\label{sec:method}

The effective use of code generation tools requires a strict regression
testing regime.  For example, an early
optimization campaign on the 1X+3V predecessor of this
code~\cite{vmlpaper} committed fusion
and staging as three ungated changes, passed its physics spot-check --- a
Landau damping rate fit, which measures only the field envelope --- and
lost $87\%$ of the mass in the system.  Every change reported above
therefore names one of four exactness classes before it is committed, and the
gate asserts that class against committed reference values on both the CPU and
the GPU build:
\begin{itemize}
\item \textbf{Bit-identical}: the run's output equals a committed
reference byte for byte; for example, a run with the poloidal plane cut
into pieces against the same run on the whole plane, and the default
path after a new option that is off by default is added.
\item \textbf{Exact degeneration}: algebraic reduction to a verified
operator in a parameter limit; for example, uniform $B$ removes the
mirror force and the drift.
\item \textbf{Machine precision, with the mechanism}: a stated bound and
the reordering that produces it; for example, the row axis at
$2.7\times10^{-12}$, through the cancellation in $\delta n$
(Sec.~\ref{sec:rows}).
\item \textbf{Physics rate}: a reference computed independently of the
code and at its own quadrature, within a stated tolerance; for example,
the slab-ITG growth rate~\cite{coppi1967} and Landau damping with the
mirror force.
\end{itemize}

Restructuring an early monolithic version of
the solver, one C++ file and its six headers with 13,348 lines, into the
KINTORA library, whose objects follow PETSc's life cycle (Create,
SetFromOptions, SetUp, View, Destroy), took 46 commits that altered no
arithmetic, each held to a byte comparison against the previous binary.
The library has since grown to 50 C++ files and 41 headers with 42,685
lines.  This regime, rather than trust in the assistant, is what made
delegating the implementation safe.  It caught a composite-step
order defect the E$\times$B term introduced, order $1.38$ in $\Delta t$ against
the design order 2; a midpoint density predictor before the field
solve restores the order to $2.59$.  \iflong Appendix~\ref{app:gates} reports three
further incidents, among them a term that every degeneration gate is
blind to.\fi

\section{Evaluation}
\label{sec:eval}

\subsection{Platform and configuration}
\label{sec:platform}

We measure two series: a strong series over the $\mu$ row axis
(Table~\ref{tab:rows}) and an accuracy-matched weak series
(Table~\ref{tab:weak}), and, around them, whether the ragged halos save
time over a worst-case exchange (Sec.~\ref{sec:raggedtime}), what each
optimization is worth, and where point C's kernels sit on the
roofline.  Both series run on Perlmutter's A100 nodes, four to a
node on Slingshot-11 and one rank per GPU\iflong; the rank-invariance
sweep runs on its CPU nodes\fi.  The physics is the full drift-kinetic
operator chain with temperature- and density-gradient profiles, $P_2$
elements and the shaped plane.  The shaped plane's operators are verified in the companion
paper~\cite{companion}: on the shaped plane the polarization solve
converges at the circle's order on a manufactured solution ($P_2$ order
$2.98$, against $2.99$ on the circle), the field-line trace converges at
fourth order, and the drift reproduces a linear state to roundoff.  At
points A and B a step on the shaped plane costs within $5\%$ of one on
the circular plane at matched sizes ($2.8\%$ less at A, $4.7\%$
more at B).

The strong series runs on point A's level-4 plane,
$N_{\mathrm{pol}} = 8{,}449$ poloidal nodes, $N_\varphi = 16$ wedges over
48 toroidal stations, the $48\times24$ velocity grid (12 uniform $\vpar$ segments of degree 3, and $\mu$
rows at the GLL nodes of 8 uniform segments of degree 2 in $w$) and
$\Delta t = 0.2$.  The weak series is three refinements, points A, B and
C, at the meshes, wedges, velocity grids and time steps that
Table~\ref{tab:weak} lists.
From one point to the next every spatial dimension and the time step are
refined $2\times$ and each velocity dimension about $\sqrt2$, $16\times$
in degrees of freedom, because the resolution needed in velocity space is
usually relatively modest.  The weak series' $\vpar$ segments are of degree 3 and
its $\mu$ rows are the GLL nodes of uniform segments of degree 3 in $w$ ($= v_\perp$ at $B_0$), not the
strong series' degree 2.

Times are per step and hot: a few steps warm up in their own log
stage and the time reported is the next stage's PETSc
\texttt{-log\_view} event time, wall clock on the host, divided by its
step count.  The tables compare the stages of the step, not run totals,
which include setup.  The runs use \texttt{-log\_view\_gpu\_time},
which times PETSc's own GPU operations with device events and
synchronizes after each; the synchronization makes the field stage,
whose solve is PETSc's, $4$--$6\%$ slower than without the option.  The first runs of points A and B of
Table~\ref{tab:weak} omitted the option; their field-stage entries are
from reruns with it, and the option moves no other stage by more than $3\%$.

\subsection{Rank and node invariance}

On Perlmutter CPU nodes a linear toroidal run at $N_\varphi = 16$ was
repeated at $P_\varphi = 1, 2, 4, 8$ and $16$ ranks.  The peak parallel
field is identical at every rank count, and the fitted mode growth rate
agrees to a worst relative deviation of $7.6\times10^{-13}$ against a
$10^{-9}$ gate tolerance.  Eight ranks on
one node and four plus four ranks across two nodes produce
bit-identical results.  Nothing in
the state path sums across toroidal ranks, and the diagnostics reduce
at their own roundoff, which is why the invariance is exact rather than
to tolerance.  The row axis adds the one reduction the state path has,
and its invariance is $2.7\times10^{-12}$ for the reason given in
Sec.~\ref{sec:rows}.

\subsection{The ragged exchange in time}
\label{sec:raggedtime}

Every channel of the weak series has depth one by design
(Sec.~\ref{sec:weak}), so the series cannot show what the ragged depths
save.  We timed instead the circular plane at refinement level 3 on four
toroidal ranks of one node at a fixed $\Delta t = 0.2$, where the depth
grows in whole wedges with $N_\varphi$: every channel has depth one at 8
and 16 wedges, and at 32 and 64 the one-sided worst case moves
$1.85\times$ the bytes of the ragged exchange.  At 32 and 64 wedges we ran
both exchanges in one allocation, bit-identical in their physics.  With
the ragged depths the stream's MPI wait is $1.8\times$ to $2.2\times$
shorter, but the step time does not change beyond the $3\%$ run-to-run
spread of a repeated point.  The likely reason is that four ranks on one
node put the toroidal exchange off the critical path, and the saving
should reach the step where the exchange crosses nodes and the wait is a
larger share of the step.  At this size the ragged depths save volume,
not time: each half-stream's halo is $13.5\%$ of a rank's owned state at
32 wedges and $10.2\%$ at 64, against $25.0\%$ and $18.8\%$ for the
one-sided worst case, a $1.85\times$ saving in the bytes on the network
and in the staging buffers that hold them.

\subsection{Strong scaling over the row axis}
\label{sec:rowscale}

Table~\ref{tab:rows} is a strong-scaling series over $\mu$ row blocks at
four fixed toroidal ranks, one to twelve blocks on 4 to 48 GPUs, on the
level-4 shaped plane with 16 wedges and the $48\times24$ velocity grid,
$4.67\times10^{8}$ phase-space DOF.  With the solves divided among the
row ranks the step's speedup reaches $7.4\times$ at twelve blocks, two $\mu$ rows per GPU, at $61\%$
efficiency.  The velocity kick divides with the $\mu$ rows almost
exactly; the drift, E$\times$B and stream stages lose $14$--$40\%$ by
twelve blocks, where a rank holds two rows, and the field stage, whose
solve does not divide, is more than half of the loss.

The field stage's per-$\mu$-row work (the moment and the predictor)
divides with the row blocks; its polarization solve does not.
When every row rank solves all twelve of its toroidal rank's stations,
the solve stays at $0.76$\,ms.  When the twelve are divided among the row
ranks, at twelve blocks, one station per
rank, the solve still takes $0.53$\,ms, the latency of a direct solve
on one plane.  Dividing the stations also requires gathering the solved
planes to every row rank, a cost timed outside the solve event, so the
field stage with the stations divided stays within $5\%$ of the field
stage in which every rank solves them all.

\begin{table}[t]
\centering
\caption{Strong scaling of the whole step, the field stage and the
polarization solve over $\mu$ row blocks at four fixed toroidal ranks
(ms per step).  Every rank: every row rank solves all of its toroidal
rank's stations.  Divided (production mode): those stations are divided among the row
ranks, and the solved planes are gathered to all of them.  The
every-rank runs were a separate allocation on a later build.}
\label{tab:rows}
\scriptsize
\setlength{\tabcolsep}{2.5pt}
\begin{tabular}{rrrrrrrrr}
\hline
row & & \multicolumn{3}{c}{every rank} & \multicolumn{4}{c}{divided} \\
blocks & GPUs & step & field & solve & step & field & solve & eff. \\
\hline
 1 &  4 & \multicolumn{3}{c}{(same computation)} & 79.23 & 14.32 & 0.76 & 100\% \\
 2 &  8 & 42.10 & 8.29 & 0.76 & 39.31 &  8.12 & 0.65 & 101\% \\
 4 & 16 & 23.36 & 5.21 & 0.76 & 22.53 &  5.23 & 0.59 &  88\% \\
 8 & 32 & 13.92 & 3.81 & 0.77 & 13.38 &  3.86 & 0.55 &  74\% \\
12 & 48 & 11.11 & 3.42 & 0.77 & 10.78 &  3.59 & 0.53 &  61\% \\
\hline
\end{tabular}
\end{table}

\iflong
\subsection{The plane-count series}

Table~\ref{tab:planes} holds the rank count at four and walks
$N_\varphi$ over 8, 16, 32 and 64 at fixed $\Delta t$ and three poloidal
refinements, where the halo width grows with the plane count.  The depth
grows only in steps, when the fastest foot passes another wedge, so
raising $N_\varphi$ buys resolution and not communication headroom,
which is the ceiling of Sec.~\ref{sec:ceiling} measured from the other
side.  Every run reports its own exchange volume against the one-sided
worst case, and the measured ratio is $1.000$ at 8 and 16 planes and
$1.846$ at 32 and 64, which is the depth arithmetic of
Sec.~\ref{sec:ragged} on the machine.  Note that the ghost event of the
table is not the exchange time: with the requests bound to the device
staging (the Artifact Description) the interior stream kernel is queued in
front of the unpack, and the event measures both.

\begin{table}[t]
\centering
\caption{The plane-count series at four toroidal ranks (A100, 100
steps).  Times in seconds, depth in wedges, volume per half-stream in MB
against the one-sided worst case.}
\label{tab:planes}
\footnotesize
\setlength{\tabcolsep}{4pt}
\begin{tabular}{rrrrrrr}
\hline
$N_\varphi$ & step & ghost & depth & ragged & worst case & ratio \\
\hline
 8 &  3.59 & 0.77 & 1 & 57.40 &  57.40 & 1.000 \\
16 &  5.76 & 1.72 & 1 & 57.40 &  57.40 & 1.000 \\
32 &  9.57 & 2.40 & 2 & 62.18 & 114.80 & 1.846 \\
64 & 18.04 & 4.81 & 3 & 93.28 & 172.20 & 1.846 \\
\hline
\end{tabular}
\end{table}
\fi

\subsection{Accuracy-matched weak scaling}
\label{sec:weak}

Table~\ref{tab:weak} is a weak series on the shaped plane of
Fig.~\ref{fig:torus} that refines all five phase-space dimensions at
once.  It is accuracy-matched in that each refinement raises every
dimension from the point before (Sec.~\ref{sec:platform}) rather than
trading one against another, and the per-GPU load is held to within
$8\%$ in degrees of freedom.  Halving $\Delta t$ as the wedges
double holds the largest shift at the same fraction of a wedge, so every
channel has depth one at every series point and the ragged and worst-case
volumes coincide: this series is not where the ragged exchange pays.  Point A is MPI serial, point
B adds the toroidal and row axes, and point C adds the plane axis, two
pieces of the plane on four toroidal ranks and 32 row blocks, the fastest
of the four rank shapes we tried at 256 GPUs.

We measured $58.8$, $78.8$ and $103.0$\,ms per step at A, B and C, a
weak-scaling efficiency (the one-GPU step time over the point's) of $75\%$ on 16
GPUs and $57\%$ on 256.  Of the
$44.2$\,ms the step grows from A to C, the stream accounts for $35.4$.
Its apply grows from $8.5$ to $24.1$\,ms, and its ghost event
(Table~\ref{tab:weak}) from $12.8$ to $32.5$.  The two row collectives, the density
all-reduce and the diagnostic reduction, grow from $0.02$ to $13.0$\,ms,
and that is waiting time: the stage times of Table~\ref{tab:weak} are
maxima over ranks, and at C they sum to $11\%$ more than the step.  The field stage
grows from $2.5$ to $9.5$\,ms while its Poisson solve stays under
$1.3$\,ms (the direct solve of Sec.~\ref{sec:field}) and its
flux-surface average grows from $0.17$ to $1.1$\,ms.  The velocity
kick, the one stage with no plane-sized table, moves by at most $13\%$
over the series.

The apply and the field stage grow because of a defect in the series'
design, which accounts for part of the series' efficiency loss.  The
series holds the state per rank fixed by dividing the velocity grid over more row ranks,
so a point-C rank holds $7.6\times$ more plane and $8\times$ fewer
channels than point A's.  At that fixed state the stream apply row of
Table~\ref{tab:weak}, which holds no communication, nearly triples, and
at point C the apply reads
its stencil values from DRAM (Sec.~\ref{sec:roofline}),
losing the reuse between neighbouring nodes that a smaller per-rank
plane can keep in cache.  The
field stage's work outside its solve,
the gather of its right-hand side, the flux-surface average and the
field's gradients, is on the plane and done in full by every row rank, so
it grows with the plane as well.  We have not yet tried blocking the
apply over the plane, which should recover the reuse with no change to
the decomposition, or dividing the field stage's plane-sized work among
the row ranks as its solves are.

\begin{table}[t]
\centering
\caption{The accuracy-matched weak series on the shaped plane (A100, one
rank per GPU; ms per step over 100 hot steps at A and B and 25 at C,
after a five-step warm-up).  Ranks are toroidal $\times$ plane $\times$
$\mu$ blocks.  Stage times are the maximum over ranks of the PETSc event.
The stream ghost row holds the stream's exchanges and the gather of its
state, and also most of the interior apply kernel, which runs while the
exchange is waited on; the stream apply row holds the rest of the apply.
The field stage includes the Poisson solve.  The mass drift is the
maximum of $|m(t) - m(0)|/m(0)$ over each run's 105, 105 and 30 steps.}
\label{tab:weak}
\footnotesize
\setlength{\tabcolsep}{4pt}
\begin{tabular}{lrrr}
\hline
 & A & B & C \\
\hline
GPUs & 1 & 16 & 256 \\
ranks & $1\times1\times1$ & $2\times1\times8$ & $4\times2\times32$ \\
poloidal nodes $N_{\mathrm{pol}}$ & $8{,}449$ & $33{,}281$ & $132{,}097$ \\
wedges (stations) & 4 (12) & 8 (24) & 16 (48) \\
$\vpar$ nodes $\times$ $\mu$ rows & $32\times32$ & $48\times40$ & $64\times64$ \\
$\Delta t$ ($a/c_s$) & 0.2 & 0.1 & 0.05 \\
phase-space DOF & $1.04\times10^{8}$ & $1.53\times10^{9}$ & $2.60\times10^{10}$ \\
\hline
stream apply & 8.45 & 19.64 & 24.11 \\
stream ghost & 12.83 & 23.30 & 32.54 \\
velocity kick & 18.71 & 16.26 & 17.38 \\
drift & 7.28 & 6.38 & 9.12 \\
E$\times$B & 6.02 & 5.78 & 8.52 \\
field stage (Poisson solve) & 2.53 (0.75) & 4.61 (0.80) & 9.45 (1.22) \\
row collectives & 0.02 & 1.52 & 12.98 \\
\hline
step, ms & 58.80 & 78.77 & 102.99 \\
efficiency & 100\% & 74.6\% & 57.1\% \\
mass drift & $2.1\times10^{-5}$ & $2.9\times10^{-6}$ & $5.9\times10^{-7}$ \\
\hline
\end{tabular}
\end{table}

\subsection{What each optimization is worth}
\label{sec:ablation}

Table~\ref{tab:ablation} measures the named optimizations of the design
at point C of the weak series, 25 hot steps per case; each case toggles
one option from the production configuration of Table~\ref{tab:weak}.
The first four cases share one allocation with a production case, which
repeats the series' point C at $103.2$\,ms per step against $102.99$.  Two cases
are measured elsewhere and quoted against their own production case: the
divided solves in a second allocation, whose production case repeats at
$105.6$\,ms, and the E$\times$B term as a
semi-Lagrangian stage against
the explicit Eulerian source at point B, because the Eulerian form
differences $f$ on an in-plane stencil whose halo between pieces of the
plane is not implemented, and so cannot run under the plane split of
point C.  The in-plane overlap of
Sec.~\ref{sec:overlap} has no off switch and is not ablated.

\begin{table}[t]
\centering
\caption{What each named optimization is worth at point C (256 GPUs,
shaped plane, $4\times2\times32$; ms per step over 25 hot steps).
``Off'' is the production configuration with that one option reversed;
``on'' is production; $\times$ is the ratio of the two step times.}
\label{tab:ablation}
\scriptsize
\setlength{\tabcolsep}{2.5pt}
\begin{tabular}{llccccc}
\hline
 & & \multicolumn{2}{c}{stage ms} & \multicolumn{2}{c}{step ms} & \\
optimization & stage & on & off & on & off & $\times$ \\
\hline
device-direct halo MPI & ghost & 32.5 & 150.7 & 103.2 & 242.0 & 2.35 \\
unrolled stream apply & apply & 24.1 & 54.3 & 103.2 & 164.7 & 1.60 \\
static drift foot table & drift & 9.2 & 42.5 & 103.2 & 137.9 & 1.34 \\
direct solve, not GAMG & field & 9.5 & 58.8 & 103.2 & 152.5 & 1.48 \\
\hline
solves divided over the rows & field & 9.5 & 13.5 & 105.6 & 109.6 & 1.04 \\
\hline
E$\times$B as an SL stage (at B) & E$\times$B & 5.4 & 79.5 & 76.3 & 150.2 & 1.97 \\
\hline
\end{tabular}
\end{table}

This data shows that the two optimizations worth most at C are the
device-direct exchange and the unrolled stream apply.  Without device-direct MPI (the first row) the drift
and E$\times$B stages slow as well ($9.2 \to 28.8$ and $8.6 \to 15.5$\,ms), because their
in-plane exchange takes the same path.  On the field path the direct
factorization is $6.2\times$ faster in the field stage than CG with
algebraic multigrid (PETSc's GAMG) on a plane of $132{,}097$ unknowns.  Dividing a toroidal rank's stations among its row ranks, rather than
having every row rank solve all of them, saves $4.0$\,ms of the field
stage: the solve falls by $4.4$\,ms, from $5.6$ to $1.2$, and gathering
the solved field to every row rank spends $0.4$\,ms of that.  At point B
the semi-Lagrangian E$\times$B stage takes $5.4$\,ms, against
$79.5$\,ms for the explicit Eulerian source it replaced.

\subsection{Where each stage sits on the roofline}
\label{sec:roofline}

\begin{table}[t]
\centering
\caption{The dominant kernel of each stage of a rank of point C's shape
on one A100.  GB/s is DRAM throughput; DRAM, L2 and L1 are percents of each level's sustained throughput (for a
cache, of its busiest sub-unit); SM is the percent of the busiest unit
in the SM; occupancy is active warps of the 64 an SM holds; waves is the
grid in units of what the device runs at once.  Bold marks the roof each
kernel runs against.}
\label{tab:roofline}
\footnotesize
\setlength{\tabcolsep}{2.5pt}
\begin{tabular}{lrrrrrrrr}
\hline
stage, kernel & GB/s & DRAM & L2 & L1 & SM & occ. & waves & regs \\
\hline
field, density gather & 1398 & \textbf{90\%} & 74\% & 20\% &  9\% & 74\% &   4.9 & 40 \\
velocity moment       & 1275 & \textbf{82\%} & 68\% & 18\% & 43\% & 94\% &   1.0 & 28 \\
drift apply           &  894 & 57\% & 72\% & \textbf{84\%} & 28\% & 54\% &  10.2 & 56 \\
velocity kick         &  267 & 17\% & 36\% & 43\% & 43\% & 24\% &   4.0 & 96 \\
stream apply          & 1075 & \textbf{69\%} & \textbf{77\%} & 62\% & 10\% & 43\% & 127.9 & 66 \\
\hline
\end{tabular}
\end{table}

The degree to which a kernel saturates the hardware indicates where to
look for code optimization.  To that end, Table~\ref{tab:roofline}
profiles the dominant kernel of each stage of a rank of point C's shape with Nsight
Compute, one run per stage through the profiler ranges the code emits
around its own timed stages.  Nsight serializes and replays every kernel
it measures, so a rank profiled inside the 256-GPU run would run in
isolation anyway, and we reproduce one point-C rank's load on eight GPUs
instead: point C's $132{,}097$-node plane in two pieces on four toroidal
ranks, with the 128 channels a point-C rank holds, $64$ $\vpar$ nodes by
2 $\mu$ rows.  Every array and kernel is then point C's; only the two
rows' $\mu$ values differ, since the global $\mu$ grid is those two rows.  A second run of the
same binary added the L2 and L1 metrics; its kernel times match the
first's to within $0.6\%$.

Four of the five kernels run at a roof.  The density gather and the
velocity moment run at the DRAM roof and the drift apply against the L1
roof.  The stream apply, $23\%$ of the step (Table~\ref{tab:weak}), runs
at $69\%$ of the DRAM roof, but at $0.34$ flop per byte, about a
twentieth of the ridge point, so the roof it is near is a low one.  It
moves 335 bytes per phase-space entry.  Each target entry reads 36
source values, the six $P_2$ nodes of three stations in each of two
source cells, and a 72-byte map row shared over its tile of two
channels, so an entry that reuses nothing moves 332 bytes: on a point-C
rank's piece of the plane the apply reuses nothing from cache, 335
measured against 332.  The channels of a tile share one map row, so they
must share one shift, and the shift is set by $\vpar$ alone, the same
in every $\mu$ row (Fig.~\ref{fig:ragged}).  The channels are ordered
$\mu$-fastest, so a tile is the $\mu$ rows a rank holds at one $\vpar$
node, and a point-C rank holds two.

The velocity kick, $17\%$ of the step, is the one kernel at no roof.  It
uses less than half of every bandwidth while only a quarter of the warps
an SM can keep resident are active: its 96 registers per lane let an SM
hold two of its blocks, and Nsight reports registers as its occupancy
limiter.  Low bandwidth at low occupancy means too few warps are
resident to hide the memory latency, so its register count, which sets
the occupancy, is where to look next.

\iflong
\section{The same design in a 1X+3V Vlasov--Poisson solver}
\label{sec:onex}

The primitives of Secs.~\ref{sec:ragged}--\ref{sec:method} were first
built in a 1X+3V Vlasov--Poisson solver with velocity-space AMR, whose
method \ifanon is developed in~\cite{vmlpaper}\else we develop
in~\cite{vmlpaper}\fi.  This section reports their evaluation there,
following the experiment design of Einkemmer's parallel SLDG
study~\cite{einkemmer2016}, so that the tokamak measurements of
Sec.~\ref{sec:eval} can be read against a uniform-grid baseline.

\subsection{The 1X+3V solver}
\label{sec:solver1x}

\subsubsection{Discretization and splitting}

The solver advances the Vlasov--Poisson system in one spatial and three
velocity dimensions (1X+3V),
\begin{equation}
\partial_t f + v_x \partial_x f + E(x,t)\, \partial_{v_x} f = 0, \qquad
-\partial_{xx} \phi = \rho - \rho_0,
\end{equation}
with $E=-\partial_x\phi$, on a periodic spatial interval discretized by
$N_x$ DG cells of degree $k$ ($N_b=k{+}1$ basis functions per cell) and a
3D Cartesian velocity mesh managed by PETSc's Landau
infrastructure~\cite{petsc,adamslandau} with $p$4est~\cite{p4est} AMR:
$N_c$ velocity cells with $N_{b3}$ nodal DOFs each, $N_v=N_c N_{b3}$
velocity DOFs total.  Strang splitting advances one step as
\begin{equation}
\label{eq:strang1x}
X\!\left(\tfrac{\Delta t}{2}\right) \;\rightarrow\;
\text{field solve} \;\rightarrow\;
V(\Delta t) \;\rightarrow\;
X\!\left(\tfrac{\Delta t}{2}\right),
\end{equation}
where $X$ advects in $x$ (each velocity DOF at its own constant speed
$v_{x,iv}$) and $V$ advects in velocity space (the electric force),
including the hybrid conforming/non-conforming AMR sweep of~\cite{vmlpaper}.

The $X$ sweep is the communication-bearing phase and the subject of
Secs.~\ref{sec:ragged} and~\ref{sec:fusion}.  For velocity DOF $iv$ the exact
SLDG update is a two-cell stencil.  Writing the shift as
$s_{iv} = v_{x,iv}\,\Delta t/(2h_x) = n_{iv} + \alpha_{iv}$ with integer
$n_{iv}$ and fractional $\alpha_{iv}\in[0,1)$, the new coefficients in cell
$c$ are
\begin{equation}
\label{eq:xupdate1x}
f^{+}_{c} \;=\; A(\alpha_{iv})\, f_{c-n_{iv}} \;+\; B(\alpha_{iv})\,
f_{c-n_{iv}-1},
\end{equation}
where $A,B$ are $N_b\times N_b$ overlap (projection) matrices precomputed
once per velocity DOF~\cite{einkemmer,vmlpaper}.  Mass conservation is
algebraically exact, because $A+B$ is a partition of unity.

\subsubsection{Parallel structure and baseline profile}

The spatial dimension is MPI-decomposed in 1D, and velocity space is not
decomposed.  Each rank stores the full velocity space of its local
spatial cells; the velocity mesh and its metadata are common to all
ranks, and the distribution-function data is partitioned in $x$, not
duplicated.  The only recurring communication is therefore (i) the
$X$-sweep halo exchange, (ii) a reduction for the charge density
$\rho(x)=\int f\,\dd^3v$, and (iii) the field solve.  All phase-space
state is device-resident (Kokkos~\cite{kokkos}).  The distribution $f$,
its cell-major transpose for the sweep, and the halo-padded working
array live on the GPU across the entire run, and hosts touch only MPI
staging buffers.

Two properties frame the optimization problem.  First, the halo is
\emph{deep}.  Stability does not limit $\Delta t$, so production steps
have $s_{\max}=v_{\max}\Delta t/(2h_x)$ of several cells, and the
distributed array is built with stencil width
$s_w = \lceil s_{\max}\rceil + 2$.  Second, the halo is \emph{oversized}
by default.  A fixed-width exchange sends $s_w$ cells of \emph{every}
velocity DOF in \emph{both} directions, although DOF $iv$ reads at most
$n_{iv}{+}1$ cells on its single upwind side, and on a velocity mesh
refined at low $|v|$, the physically interesting region, most DOFs have
small $n_{iv}$.

\paragraph{The ragged exchange, measured there}
Per-DOF ragged depths give volume
$\sum_{iv} \big(d_L(iv)+d_R(iv)\big) N_b$.  In the reference configuration
(Q3 velocity space, $4^3$ cells + 1 AMR level, $R=6$, $N_x=64$,
$\Delta t=0.1$) the baseline sends $69{,}120$ elements per exchange and
the ragged exchange $14{,}400+11{,}712=26{,}112$, a ratio of $2.6\times$.
The gain is the ratio of the worst-case to the mean per-DOF depth, so it
grows with velocity-space AMR.  Refinement concentrates DOFs at low
$|v|$, where depths are small, while the worst case is pinned by
$v_{\max}$.  On this configuration the gain, $2.6\times$, exceeds the
uniform-mesh estimate of about $2\times$.  The velocity meshes that
improve the physics also shrink the messages.  Not decomposing velocity
is a design point, not a limitation.  With velocity-space AMR the
velocity mesh stays modest.  The refined $(v_\parallel,v_\perp)$ and 3V
meshes of interest are $O(10^2$--$10^3)$ cells.  With Q3 elements (16
DOFs per 2V cell, 64 per 3V cell) and five species, the velocity space
at one spatial node holds $0.26$ to $2.6$\,MB in 3V and $64$\,KB to
$0.64$\,MB in 2V, at 8 bytes per DOF, which fits the on-chip memory of
current GPUs.  A thread-block cluster can then process one spatial
node's velocity space cooperatively.  Appendix~\ref{app:onchip}
works this through for B200-class hardware.  On host-only nodes the
persistent requests alone cut halo-exchange time $36\%$ at 8 ranks.  On
A100 nodes the combined ragged path holds the per-exchange cost to
$\sim$1\,ms, and unlike the fixed-width exchange it \emph{shrinks} when
crossing from one node to two (Sec.~\ref{sec:eval1x}).

\subsection{Exact operator fusion}
\label{sec:fusion}

\subsubsection{Two projected half-steps are not one full step}

In~\eqref{eq:strang1x}, the trailing $X(\Delta t/2)$ of step $n$ and the
leading $X(\Delta t/2)$ of step $n{+}1$ are adjacent with no operator
between them.  One can replace the pair by a single SLDG sweep
$X(\Delta t)$ built from overlap matrices at the doubled shift, which
halves the exchanges, and an earlier version of this code did exactly
that.  But the
SLDG update is ``shift exactly, then $L^2$-project onto the DG space,''
$X_h(s) = P_h T(s)$, and the intermediate projection does not commute with
the shift:
\begin{equation}
P_h T(s) P_h T(s) \;\ne\; P_h T(2s)
\end{equation}
in general.  The $P^0$ (finite-volume) case shows the difference.  Two
projected half-cell shifts yield the averaging stencil
$(f_{j-2}+2f_{j-1}+f_j)/4$, while one full-cell shift gives $f_{j-1}$
exactly.  The single-projection full step is a \emph{valid but different}
scheme, less diffusive by one projection per step, and measured
against the unfused trajectory it drifts at $O(\Delta t^2)$ (a relative
deviation of $5\times10^{-5}$ at $t=2$ in our earlier experiments).  That drift fails
an acceptance criterion of reproducing verified physics trajectories.

\subsubsection{Fusing by matrix composition}

The correct fusion composes the discrete operators themselves.  Applying
\eqref{eq:xupdate1x} twice and expanding,
\begin{equation}
\label{eq:fused}
f^{++}_{c} = \underbrace{AA}_{C_0} f_{c-2n}
           + \underbrace{(AB+BA)}_{C_1} f_{c-2n-1}
           + \underbrace{BB}_{C_2} f_{c-2n-2},
\end{equation}
a three-cell stencil with shift $2n$ and matrices precomputed once per
velocity DOF at setup.  Equation~\eqref{eq:fused} is an algebraic
identity.  The fused sweep differs from the sequential pair only by
floating-point reassociation, conserves mass to roundoff (products of
partition-of-unity pairs), and needs halo depths $2n{+}2$ / $-2n$.  One
exchange carries what two did, at the same total volume but half the
message count, pack/unpack passes, and kernel launches.  The first and
last steps of a run keep bare half-steps.  A pending-half-step flag in
the time loop selects the bare or the fused leading sweep, and per-step
moment diagnostics on the unmaterialized state are exact to roundoff
because the $X$-transfer conserves every per-DOF spatial integral.

Measured against the unfused optimized code on A100s, fusion cut total
transport-sweep time $30$--$39\%$ at 4 ranks and $16\%$ at 8.  Ten-step
trajectories agree to $\sim\!10^{-15}$ relative, and 100-step
verification runs match the committed physics reference data of
\cite{vmlpaper} to $\sim\!10^{-12}$, the same level as the unfused code.

\subsection{A replicated direct field solve}
\label{sec:poisson}

After the transport and reduction phases were optimized, the field solve
dominated all non-transport cost on GPU nodes, and it \emph{grew} from 4
to 8 ranks: $0.13$ against $0.72$\,s per 20-step run of the large
configuration.  The system is
small, one global 1D periodic Poisson equation with $N_x\le 1024$
unknowns and an 8\,KB right-hand side, so a distributed
conjugate-gradient/AMG solve at $10^{-14}$ tolerance is pure inter-node
latency.  Every iteration is two allreduces plus a multigrid cycle of
small messages.

The replacement gathers the right-hand side (\texttt{MPI\_Allgatherv},
one small collective) and solves redundantly on every rank.  Grounding
one node of the periodic ring reduces the system to a pure tridiagonal
chain (the wrap couplings pass through the grounded node), which the
Thomas algorithm solves, and the result is shifted to the mean-zero
representative that the Krylov path's null-space projection would
return.  The solve is exact to machine precision, \emph{more} accurate
than the iterative solver it replaces, and its cost is $O(N_x)$ flops
per rank plus one latency term.  The field phase falls from $0.75$\,s to
$9.1$\,ms per 100-step run of the reference case at 4 ranks, and from
$0.13$\,s to $3.7$\,ms and from $0.72$\,s to $6.9$\,ms per 20-step run
of the large configuration at 4 and 8 ranks, and the growth with rank
count is gone.

Two remarks generalize the point.  First, the fix is fewer messages, not
a GPU port.  A system of $10^3$ unknowns does not need an A100.  Second,
the implementation sits behind a \emph{batched} field-solve interface.
The tokamak solver solves one 2D perpendicular polarization system per
poloidal plane, a batch of independent solves, and the redundant 1D path
is that interface's batch~$=1$ backend.

\subsection{Evaluation}
\label{sec:eval1x}

\subsubsection{Platforms and configurations}

\textbf{Perlmutter (NERSC).}  GPU nodes with 4$\times$NVIDIA A100-80GB
and Slingshot-11, the PETSc/Kokkos CUDA backend (sm\_80), Cray MPICH,
and one rank per GPU.  \textbf{Development gate platform.}  Apple
M-series with the Kokkos Serial backend, under AddressSanitizer for the
gates and as an optimized build for host timings.

\textbf{Model problem.}  We adopt the experiment design of Einkemmer's
parallel SLDG study~\cite{einkemmer2016}, adjusted from 1X+1V on a CPU
cluster to 1X+3V on GPU nodes.  The physics case is strong Landau damping
on $[0,4\pi]\times[-6,6]^3$ with
$f_0=(2\pi)^{-3/2}e^{-|\mathbf{v}|^2/2}\,(1+\tfrac12\cos\tfrac{x}{2})$.
Communication is parameterized by the CFL number
$C=v_{\max}\Delta t/h_x$ (per-DOF $C_{iv}=|v_{x,iv}|\Delta t/h_x$).
Dimensionality forces one adaptation of~\cite{einkemmer2016}.  His
worst-case weak scaling (fixed cells per rank, fixed $\Delta t$) grows
$C$ as $P^{1/d}$ under a $d$-dimensional phase-space decomposition; in
his 2D and 4D advection studies, $C\!\sim\!91$ at 8192 cores in 2D but
only $\sim\!12$ in 4D.  It rests on the assumption $C<n_P$ (cells per
rank per dimension), which keeps the exchange nearest-neighbor and the
buffer-to-local-data ratio $C/n_P$ bounded.  Our spatial decomposition
is one-dimensional, so the same shape grows $C$ \emph{linearly} in $P$
and breaks $C<n_P$ at modest scale.  We therefore run four series.  Weak
scaling comes first, because for kinetic equations resolution must grow
with the machine to capture filamentation, which makes weak scaling the
operative metric~\cite{einkemmer2022}.  (a) \emph{Accuracy-matched weak
scaling}, the primary series, holds $\Delta t \propto h_x$, so $C$ and
the communication fraction are constant, the shape a production
simulation uses.  (b) \emph{Strong scaling} runs at moderate fixed
$C\approx5$.  (c) A \emph{worst-case sweep in $C$} at fixed machine size
and fixed $\Delta t$, up to $C\approx n_P$, reports the ragged-halo gain
as a function of $C$, the regime where halo depth dominates.  (d)
\emph{Pure-advection scaling in 2X} follows~\cite{einkemmer2016}, whose
scaling studies are advection tests, with a 2X$+$3V (and 2X$+$2V)
constant-coefficient advection benchmark on a two-dimensional spatial
decomposition: two sweep directions, four neighbors per rank, and the
process topology the tokamak solver uses.  It exercises the ragged
exchange beyond the 1D ring without requiring the field solve.  All four
series run on the model problem with a Q3, $4^3{+}$1-AMR velocity mesh
(120 cells, $N_v=7{,}680$ velocity DOFs).  The worst-case-depth
one-sided exchange of~\cite{einkemmer2016} is implemented behind a
runtime flag in the same code, with identical kernels and staging and
every per-DOF depth padded to its direction's maximum, so the comparison
isolates message volume.  The flag reproduces the ragged trajectory
bit-identically, because the padding is written and never read.  The
2X$+$2V advection driver shares the sweep and exchange machinery of
Secs.~\ref{sec:ragged}--\ref{sec:fusion} (the transverse spatial DOFs
fold into the per-channel block), runs on a $p_x\times p_y$ rank grid
with a device transpose between the dimension-by-dimension sweeps, and
is gated by its exact solution.  Time integration is exact for constant
coefficients, the measured spatial order is $3.0$ (Q2) and $4.2$ (Q3),
and results are invariant across process topologies.  Every campaign
run asserts mass conservation to $10^{-10}$ at postprocessing, and all
runs passed.

\begin{figure}[t]
\centering
\begin{tikzpicture}[font=\footnotesize]

  \begin{scope}
    \node[anchor=north west, font=\footnotesize\bfseries] at (-0.15,3.45) {(a)};

    \begin{scope}[shift={(0.5,0)}]
      \fill[blue!8] (1,1) rectangle (2,2);
      \draw[black!35] (0,0) grid (3,3);
      \draw[line width=1.2pt, blue!70!black] (1,1) rectangle (2,2);

      \node[font=\tiny, black!60] at (0.5,2.5) {$(0,2)$};
      \node[font=\tiny, black!60] at (2.5,2.5) {$(2,2)$};
      \node[font=\tiny, blue!70!black] at (1.5,1.72) {$(1,1)$};
      \node[font=\tiny, black!60] at (0.5,0.5) {$(0,0)$};
      \node[font=\tiny, black!60] at (2.5,0.5) {$(2,0)$};

      \draw[->, semithick, blue!70!black] (1.5,2.05) -- (1.5,2.45);
      \draw[->, semithick, blue!70!black] (1.5,0.95) -- (1.5,0.55);
      \draw[->, semithick, blue!70!black] (2.05,1.5) -- (2.45,1.5);
      \draw[->, semithick, blue!70!black] (0.95,1.5) -- (0.55,1.5);

      \draw[->, semithick, black!45, dashed]
        (3.05,0.25) .. controls (3.55,-0.30) and (-0.55,-0.30) .. (-0.05,0.25);
      \node[font=\tiny, black!45] at (1.5,-0.40) {periodic};

      \node[font=\tiny] at (1.5,3.20) {$p_x \times p_y$ ranks in $(x,y)$};
    \end{scope}

    \begin{scope}[shift={(5.6,0.42)}]
      \draw[black!55] (0,0) grid[step=0.36] (2.16,2.16);
      \draw[->, black!55] (-0.12,-0.12) -- (2.38,-0.12) node[right, font=\tiny] {$v_x$};
      \draw[->, black!55] (-0.12,-0.12) -- (-0.12,2.38) node[above, font=\tiny] {$v_y$};
      \node[font=\tiny, align=center, anchor=north] at (1.08,-0.38)
        {full $(v_x,v_y)$ grid in \emph{every}\\ spatial cell (not decomposed)};
    \end{scope}
    \draw[->, semithick, black!55] (3.75,1.5) -- (5.45,1.5);
  \end{scope}

  \begin{scope}[shift={(0,-4.35)}]
    \def\xl{1.75}
    \def\xr{5.05}
    \def\dmax{1.5}

    \node[anchor=north west, font=\footnotesize\bfseries] at (-0.15,3.30) {(b)};

    \draw[dashed, black!55] (\xl-\dmax,0.28) rectangle (\xr+\dmax,2.52);

    \fill[blue!7] (\xl,0.28) rectangle (\xr,2.52);
    \draw[blue!70!black, line width=1pt] (\xl,0.28) -- (\xl,2.52);
    \draw[blue!70!black, line width=1pt] (\xr,0.28) -- (\xr,2.52);
    \node[font=\tiny, blue!70!black, anchor=south]
      at (0.5*\xl+0.5*\xr,2.56) {owned cells of one rank};

    \foreach \yc/\dl/\dr/\lab in {
      2.28/1.5/0/{$v_x \gg 0$},
      1.76/0.5/0/{$v_x > 0$},
      1.24/0/0.5/{$v_x < 0$},
      0.72/0/1.5/{$v_x \ll 0$}}
    {
      \draw[black!65, line width=0.4pt] (\xl,\yc-0.18) rectangle (\xr,\yc+0.18);
      \ifdim\dl pt>0pt
        \fill[red!22] (\xl-\dl,\yc-0.18) rectangle (\xl,\yc+0.18);
        \draw[red!60!black, line width=0.4pt]
          (\xl-\dl,\yc-0.18) rectangle (\xl,\yc+0.18);
        \draw[->, semithick, red!60!black] (\xl-\dl+0.08,\yc) -- (\xl-0.08,\yc);
      \fi
      \ifdim\dr pt>0pt
        \fill[red!22] (\xr,\yc-0.18) rectangle (\xr+\dr,\yc+0.18);
        \draw[red!60!black, line width=0.4pt]
          (\xr,\yc-0.18) rectangle (\xr+\dr,\yc+0.18);
        \draw[<-, semithick, red!60!black] (\xr+0.08,\yc) -- (\xr+\dr-0.08,\yc);
      \fi
      \node[anchor=west, font=\tiny] at (\xr+\dmax+0.14,\yc) {\lab};
    }
    \draw[->, black!70] (\xl-\dmax,0.06) -- (\xr+\dmax,0.06)
      node[right, font=\tiny] {$x$};

    \begin{scope}[shift={(0.25,-0.85)}]
      \fill[red!22] (0,0) rectangle (0.30,0.17);
      \draw[red!60!black, line width=0.4pt] (0,0) rectangle (0.30,0.17);
      \node[anchor=west, font=\tiny] at (0.38,0.085)
        {ragged: upwind side only, depth $\propto |v_x|$};
      \draw[dashed, black!55] (0,-0.36) rectangle (0.30,-0.19);
      \node[anchor=west, font=\tiny] at (0.38,-0.275)
        {worst case: every channel padded to the maximum, both sides};
    \end{scope}
  \end{scope}

\end{tikzpicture}
\caption{2X+2V benchmark domain and the ragged halo.  \emph{(a)} The
$p_x \times p_y$ rank grid (here $3\times 3$), periodic in both
directions; the highlighted rank exchanges with four neighbours.
Velocity space is not decomposed: every spatial cell carries the full
$(v_x,v_y)$ grid.  \emph{(b)} One X-sweep, for four representative
velocity channels.  A channel's halo depth is proportional to $|v_x|$
and only its upwind side is exchanged, so the region actually sent
(solid) is a fraction of the fixed-width envelope a worst-case
formulation moves (dashed).  Sec.~\ref{sec:ragged} gives the
prefix-summed layout.}
\label{fig:ex8domain}
\end{figure}

Fig.~\ref{fig:ex8domain} sketches the domain decomposition and the ragged
halo structure.  Each velocity DOF requires a depth proportional to its
speed, so an exchange of upwind-only halos alone cuts communication
volume.

We report an \emph{effective sweep bandwidth}, the minimum streaming
traffic of a sweep (one read and one write of the distribution,
$2\times 8 N_v N_x N_b$ bytes) divided by the sweep wall time including
its exchange.  The numerator is fixed at this minimum rather than the
bytes a given exchange strategy moves, so it is insensitive to how much
traffic the ragged halo saves.  A ratio of two effective-bandwidth
numbers is exactly the wall-clock speedup between the two strategies,
not a hardware-roofline utilization figure.  This is an end-to-end
number.  Per-kernel DRAM bandwidths from Nsight Compute are reported
separately below.

\subsubsection{Phase times per optimization state}

Table~\ref{tab:cuda} summarizes the Perlmutter measurements.  Each
per-phase effect is attributable to one design section above.  The
cooperative-team charge-density reduction costs $4.7$\,ms per run, $8\%$
to $14\%$ of the transport sweep.  The ragged persistent exchange holds
the halo to about 1\,ms per exchange and \emph{improves} when crossing to
a second node: before fusion the halo event falls from $44.1$\,ms per run
at four ranks on one node to $35.4$\,ms at eight on two.  Fusion halves
the exchange count and takes a further $30\%$ off the transport sweep at
four ranks and $16\%$ at eight.  The replicated field solve reduces the
field phase to $3.7$ and $6.9$\,ms per run.  After all seven
optimizations, everything but the velocity sweep costs about $50$\,ms per
run at 8 ranks against the sweep's $0.40$\,s, so the code is again
dominated by its arithmetically dense phase.

\begin{table}[t]
\caption{Perlmutter A100 phase times in seconds per 20-step run, \emph{big}
configuration ($N_x{=}1024$; np = ranks = GPUs).  Cumulative optimization
state per row group; AdvectX = total transport sweep including exchange.}
\label{tab:cuda}
\centering
\small
\setlength{\tabcolsep}{4pt}
\begin{tabular}{llrrrr}
\toprule
Opt state & np & AdvectX & Halo & Poisson & $\rho$ \\
\midrule
\multirow{2}{*}{A--D (ragged halo)} & 4 & 0.061 & 0.044 & 0.129 & 0.0047\\
 & 8 & 0.045 & 0.035 & 0.703 & 0.0042\\
\multirow{2}{*}{A--E (+overlap)} & 4 & 0.050 & 0.041 & 0.131 & 0.0048\\
 & 8 & 0.043 & 0.037 & 0.716 & 0.0041\\
\multirow{2}{*}{A--F (+fusion)} & 4 & 0.035 & 0.028 & 0.130 & 0.0047\\
 & 8 & 0.036 & 0.031 & 0.763 & 0.0043\\
\multirow{2}{*}{A--G (+direct field)} & 4 & 0.034 & 0.027 & \textbf{0.0037} & 0.0048\\
 & 8 & 0.035 & 0.030 & \textbf{0.0069} & 0.0043\\
\bottomrule
\end{tabular}
\end{table}

\subsubsection{Ragged vs.\ worst-case exchange across the Courant range}

\begin{figure}[t]
\centering
\includegraphics[width=\linewidth]{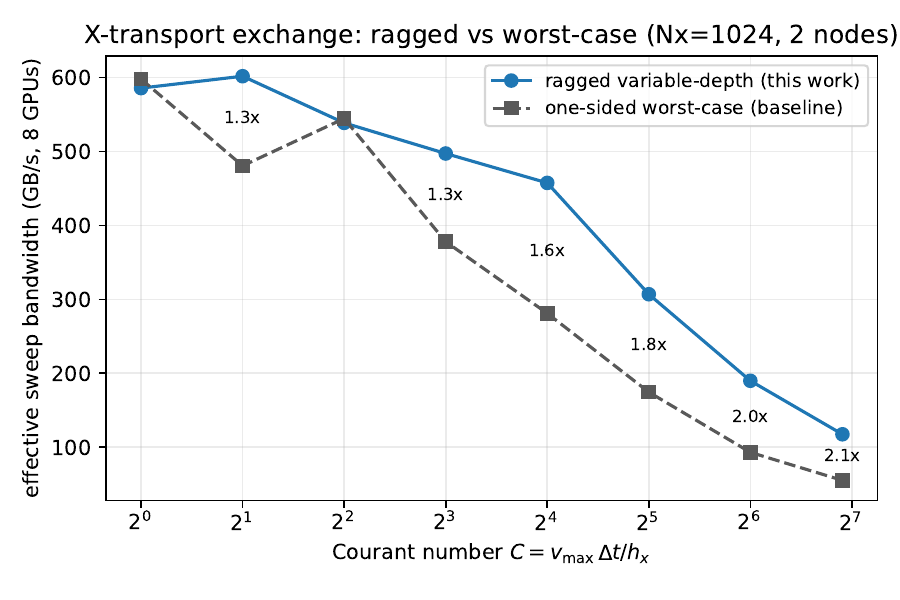}
\caption{Series (c): effective sweep bandwidth of the ragged
variable-depth exchange and the worst-case-depth
baseline~\cite{einkemmer2016} versus Courant number
($N_x{=}1024$, 2 nodes / 8 GPUs, fixed problem, $\Delta t$ varied).
Annotations give the ratio.}
\label{fig:csweep}
\end{figure}

Fig.~\ref{fig:csweep} is the direct measurement of the ragged-halo claim.
At $C=1$ every velocity DOF needs depth 0 or 1 and the two schemes send
nearly the same data, so the curves coincide.  As $C$ grows the worst
case is pinned by $v_{\max}$ while the mean depth grows with the mean
speed of the Maxwellian bulk, and the measured advantage opens
monotonically, to $1.3\times$ at $C=8$, $1.6\times$ at $C=16$,
$1.8\times$ at $C=32$, and $2.1\times$ at $C=120$, the largest $C$ the
local subdomain width admits ($C \lesssim N_x^{\mathrm{loc}} = 128$).
Large $C$ is the operating regime these methods are chosen for.
Semi-Lagrangian transport has no CFL stability limit, and for
constant-coefficient transport the time integration is exact, so the
right end of the figure is where production runs sit.

\subsubsection{Cumulative ablation ladder}

\begin{figure}[t]
\centering
\includegraphics[width=\linewidth]{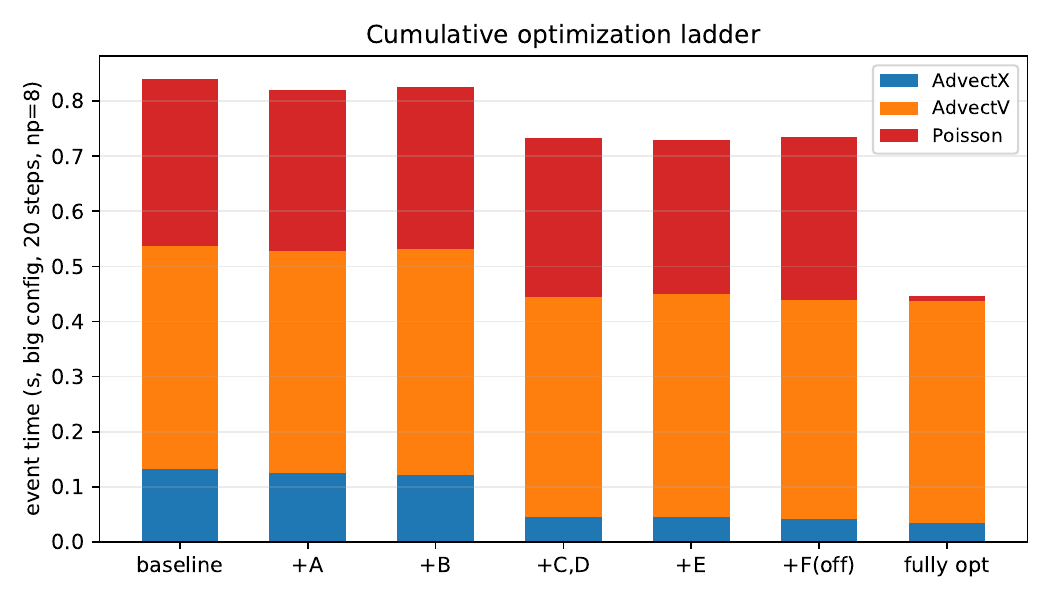}
\caption{Cumulative optimization ladder: each rung is the buildable
revision at which that optimization was committed, rebuilt from the commit
history inside the job and run on the large configuration
($N_x{=}1024$, 20 steps, 8 GPUs).}
\label{fig:ladder}
\end{figure}

Because each optimization was committed as one gated change, every rung of the
ladder is a buildable revision.  The campaign job checks each one out,
builds it, and runs the large configuration (Fig.~\ref{fig:ladder}).
The transport sweep drops from $0.132$\,s at the baseline to $0.046$\,s
when the ragged halo is committed (+C,D; $2.9\times$, almost entirely exchange
time, $0.123\rightarrow0.037$\,s) and to $0.035$\,s with fusion enabled.
The replicated field solve takes the Poisson bar from $0.30$\,s per run
to $7.4$\,ms.  The CG and multigrid solve it replaces is inter-node
latency and varies with the allocation: the run of Table~\ref{tab:cuda}
measured $0.70$ to $0.76$\,s for it at the same configuration.  The cost the optimizations target, everything
except the velocity sweep, falls from $0.434$\,s to $0.042$\,s
($10.3\times$).  The velocity sweep is unchanged at $0.40$\,s
throughout, so the optimized code spends $91\%$ of its step in its
arithmetically dense phase.

\subsubsection{Kernel bandwidths}

Per-kernel DRAM bandwidth is measured with Nsight Compute as
$\mathrm{bytes}_{\mathrm{read}}{+}\mathrm{bytes}_{\mathrm{write}}$ over
kernel duration on one A100.  The velocity sweep, the dominant kernel,
runs at $616$\,GB/s, comparable to the published single-A100 SLDG rate
of $\sim$600\,GB/s~\cite{einkemmer2022}.  The cell-major transpose that
feeds the transport sweep runs at $659$\,GB/s and the charge-density
reduction at $993$\,GB/s.  In the 2X advection driver the sweep kernel
reaches $839$\,GB/s and the transpose $1024$\,GB/s.  The one kernel
below the streaming envelope is the 1X transport apply at $260$\,GB/s,
whose stencil access pattern is layout-bound.  In the profiled run its
kernel time is $1.3$\,ms against the velocity sweep's $22.6$\,ms, so we
note it as future work rather than optimize it here.

\subsubsection{Scaling}

\begin{figure}[t]
\centering
\includegraphics[width=\linewidth]{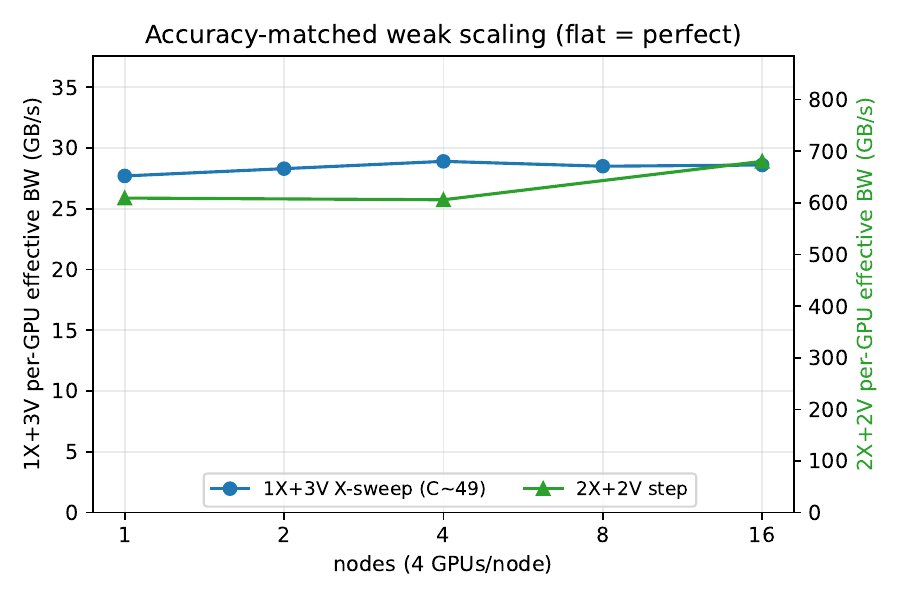}
\caption{Series (a) and (d): accuracy-matched weak scaling.  Per-GPU
effective sweep bandwidth for the 1X+3V transport sweep at $C\approx49$
(left axis) and per-GPU effective step bandwidth for the 2X+2V driver
(right axis); flat is perfect.}
\label{fig:weak}
\end{figure}

\begin{figure}[t]
\centering
\includegraphics[width=\linewidth]{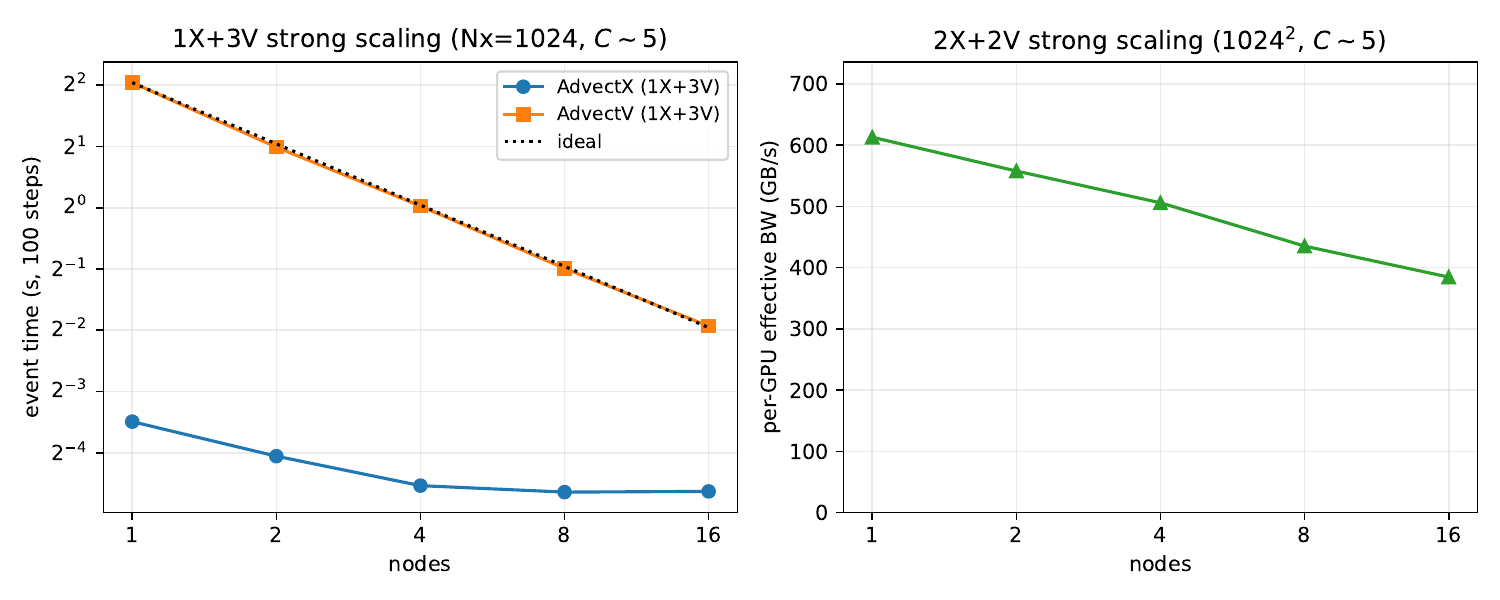}
\caption{Series (b) and (d): strong scaling.  Left: 1X+3V event times at
fixed $N_x{=}1024$, $C\approx5$; the velocity sweep follows the ideal
line while the transport sweep reaches its message-latency floor by 4
nodes.  Right: 2X+2V per-GPU effective bandwidth at fixed $1024^2$.}
\label{fig:strong}
\end{figure}

\emph{Weak scaling} (Fig.~\ref{fig:weak}) is the primary series.  With
128 spatial cells per GPU, $\Delta t\propto h_x$, and $C\approx49$ held
fixed, per-sweep transport time is constant to $3\%$ from 4 to 64 GPUs
($1.70\rightarrow1.65$\,ms per sweep as $N_x$ grows
$512\rightarrow8192$), and the per-step velocity-sweep time is constant
to $2\%$.  A one-dimensional ring keeps this series topology-simple.
The informative statement is not the flatness itself but that the
deep-halo exchange ($s_w=51$ cells) does not disturb it.  The 2X+2V
driver, whose $p_x\times p_y$ grid has four neighbors and two sweep
directions, holds $609$--$680$\,GB/s per GPU over the same range, so the
weak-scaling shape survives the two-dimensional topology the tokamak
solver uses.

\emph{Strong scaling} (Fig.~\ref{fig:strong}) at fixed $N_x{=}1024$ and
$C\approx5$ separates the two phases.  The velocity sweep is
$15.7\times$ faster on 16 nodes than on 1 ($98\%$ efficiency), because
it is compute-heavy and rank-local.  The transport sweep gains only
$2.2\times$ before flattening at 4 nodes.  With $85\%$ of its time in
the exchange, its floor is message latency, not bandwidth, which is the
component fusion halves (Sec.~\ref{sec:fusion}) and the reason exchange
count, not just volume, is a design target.  The 2X+2V driver at fixed
$1024^2$ retains $63\%$ per-GPU efficiency at 64 GPUs.  Runs beyond 16
nodes were not made.

\subsubsection{Verification results}

\begin{figure}[t]
\centering
\includegraphics[width=\linewidth]{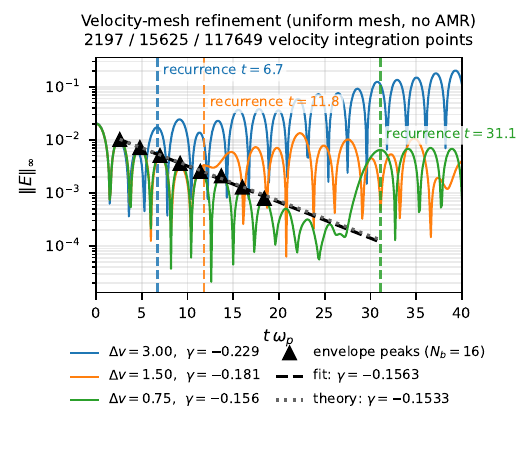}
\caption{Weak Landau damping ($\alpha=0.01$, $k=0.5$) under velocity-mesh
refinement.  Three levels at $\Delta v = 3.0$, $1.5$, $0.75$ ($N_b^3$
velocity cells of degree 3 with $N_b=4$, $8$, $16$, and $N_x=64$, $128$,
$256$).  The velocity mesh is uniform here, with AMR off, so $\Delta v$
is a single number per level.  Dashed vertical lines mark the measured recurrence onset, the
first envelope peak that exceeds its predecessor.  Markers are the
envelope peaks of the finest level and the black dashed line their
least-squares fit, as in~\cite{einkemmer2020}, drawn only over the
interval it is fitted on; the grey dotted line is the analytic rate.
The view is clipped to $t\le40$; the runs continue to $t=100$.}
\label{fig:landau}
\end{figure}

With all optimizations active, the 10-step mass drift is
$3.2\times10^{-15}$ on A100s at 1, 4, and 8 ranks.  Step-1 mass matches
the value published with~\cite{vmlpaper} to $10^{-13}$.  The 100-step
trajectories reproduce the committed reference data to $\sim\!10^{-12}$
in field maximum and $\sim\!10^{-13}$ in mass and energy on both
backends.  The fused and unfused paths, and the direct and Krylov paths,
agree to solver tolerance ($10^{-15}$, $10^{-14}$).  All 43 campaign
runs conserve mass below the $10^{-10}$ postprocessing gate.
Fig.~\ref{fig:landau} shows the physics-level check as a
velocity-refinement sequence.  Each halving of $\Delta v$ moves the
fitted damping rate toward the analytic $-0.1533$, through $-0.2288$,
$-0.1809$ and $-0.1563$, or $49\%$, $18\%$ and $2.0\%$ relative error.
All three overestimate the magnitude, so the sequence approaches the
analytic rate from one side, because an under-resolved velocity mesh
damps too fast.  The interval over which the damping is exponential lengthens
under the same refinement, with the envelope rebounding at $t=6.7$,
$11.8$ and $31.1$.  The discrete-velocity estimate $2\pi/(k\,\Delta v)$
gives $4.2$, $8.4$ and $16.8$ here.  It tracks the scaling but sits low,
because degree-3 elements resolve structure below the cell width.  A
damping fit is
therefore meaningful only inside the measured onset, which is where the
fit window ends.  Velocity space is undecomposed
(Sec.~\ref{sec:ragged}), so this sequence is a property of the
discretization and does not depend on the rank count that produced it.
\fi

\section{Related work}
\label{sec:related}

Conservative SLDG methods for kinetic equations were developed by
Crouseilles, Mehrenberger, and Sonnendr\"ucker and by
Einkemmer~\cite{crouseilles,einkemmer}.  The closest prior work is the
SLDG code line of Einkemmer et al.: distributed-memory CPU
parallelization with worst-case-depth one-sided
halos~\cite{einkemmer2016}, which is our communication baseline; GPU
Vlasov--Poisson within one node~\cite{einkemmer2020}; uniform-grid
4d/5d/6d Vlasov--Poisson on up to 1536 A100s~\cite{einkemmer2023}; and
drift-kinetic ITG on a structured cylindrical mesh on one
GPU~\cite{einkemmer2022}.  That line decomposes all phase-space
dimensions on uniform grids.  Kormann, Reuter and Rampp decompose a
six-dimensional semi-Lagrangian Vlasov--Poisson solver over all of its
dimensions with halos exchanged between neighbours, meet the time-step
restriction that decomposition imposes with a blocked one-sided
communication scheme, and hide the halo exchange behind
computation~\cite{kormann2019}; BSL6D runs that algorithm on CPU, AMD
and NVIDIA nodes from one Kokkos code base~\cite{bsl6d}.  In that line, halo depth is uniform per face at the
worst case, and depth adaptivity arises only block-wise, from which
velocity range a rank owns.  The present work differs in three ways:
per-channel ragged depths inside a decomposition that never splits
$\vpar$, unstructured extruded meshes on which the displacement is a
property of each pencil, and a second decomposition axis over the
magnetic-moment rows with its field path.  SLDG sweeps are
memory-bound, and the Einkemmer line reports an achieved bandwidth
($\sim$470\,GB/s on V100, $\sim$600\,GB/s on A100) defined as the
minimum data volume of a step, one read and one write of $f$ per
advection, over the step time~\cite{einkemmer2020,einkemmer2022}.
Table~\ref{tab:roofline} reports instead the DRAM bytes each kernel
moves, measured by Nsight Compute, so the two are not directly
comparable\iflong{}; Sec.~\ref{sec:onex} reports both for the 1X+3V
solver\fi{}.

Among tokamak codes, GYSELA~\cite{gysela} is the semi-Lagrangian
gyrokinetic code on structured field-aligned grids and the nearest
methodological relative, and XGC~\cite{xgc} handles unstructured meshes
with particles, whose communication is particle migration rather than a
halo.  Distributed GPU kinetic codes span the discretization space: DG
kinetics in Gkeyll~\cite{gkeyll}, continuum gyrokinetics on GPUs in
GENE~\cite{genegpu} and the GPU-native GX~\cite{gx}, and
particle-in-cell in WarpX~\cite{warpx}; none uses semi-Lagrangian
transport on an unstructured mesh, so their halos are stencil-width,
not $\Delta t$-dependent.  On the communication side, persistent MPI
requests are standard and our staging follows established GPU-aware MPI
practice~\cite{mvapichgpu}.  Ghost-cell expansion~\cite{dinghe2001}
deepens a halo on purpose, trading volume for fewer exchanges; here the
depth is set by $\Delta t$ and each channel's speed, and the ragged
exchange removes the volume a worst-case depth would add.  Sparse
collectives~\cite{hoefler2009}, standardized as MPI-3's neighbourhood
collectives, could carry the same exchange with per-neighbour counts;
with two toroidal neighbours per rank we use persistent point-to-point
requests bound to device staging buffers instead.  MPI-4 \emph{partitioned}
communication~\cite{mpi40,finepoints} targets multi-threaded producers
filling one buffer and is a natural future fit for the ragged
pack.\iflong{}  Operator fusion in stencil compilers fuses
\emph{kernels}.  The fusion of Sec.~\ref{sec:fusion} composes
\emph{numerical operators} with an exactness argument, which we have not
seen elsewhere for SLDG.\fi{}

\section{Limitations}
\label{sec:limits}

Limitations stated in preceding sections are collected here.
\begin{itemize}
\item \textbf{Mass conservation.}  The stream's evaluation where a field
line crosses a station and the in-plane stages' evaluation at their feet
are pointwise interpolations and do not conserve mass; the full step's
mass drift stays below $2.1\times10^{-5}$ over the series' runs
(Secs.~\ref{sec:intro} and~\ref{sec:weak}).
\item \textbf{The ragged halo's saving.}  In the timed series every
channel has depth one, so the ragged exchange moves the bytes of the
worst case.  Where the depths differ, at 32 and 64 wedges on one node, it
moves $1.85\times$ fewer bytes through $1.85\times$ smaller staging
buffers, but the step time does not change; an exchange across nodes at
those depths is not measured (Sec.~\ref{sec:raggedtime}).
\item \textbf{The weak series' design.}  Holding the state per rank
fixed by dividing the velocity grid gives a point-C rank $7.6\times$
point A's plane, so the stream apply and the field stage's plane-sized
work grow over the series; blocking the apply over the plane and
dividing that work among the row ranks are untried
(Sec.~\ref{sec:weak}).
\item \textbf{Row collectives.}  At 256 GPUs the two row collectives
hold $13.0$\,ms per step of waiting, which the design does not overlap
(Sec.~\ref{sec:weak}).
\item \textbf{The field solve.}  Each plane is solved on one GPU, so the
solve does not fall below one plane's time, $0.53$\,ms at twelve row
blocks, and dividing the stations among the row ranks adds a gather of
the solved field (Secs.~\ref{sec:field} and~\ref{sec:rowscale}).
\item \textbf{The $\mu$ row axis.}  It relies on no stage coupling the
$\mu$ rows; a collision operator would mix them and require a transpose
of the state across the axis (Sec.~\ref{sec:disc}).  The block count
must divide the number of $\mu$ rows, and the rows are partitioned by
count, not by work (Sec.~\ref{sec:rows}).
\item \textbf{Scope.}  Every timing is on Perlmutter's A100s, to 256
GPUs.  The two series, the ablation and the roofline are measured on the
shaped plane, on which the companion paper verifies the polarization
solve (the circle's order on a manufactured solution), the field-line
trace (fourth order) and the drift (a linear state to roundoff); the
ragged-exchange timing is on the circular plane
(Secs.~\ref{sec:platform} and~\ref{sec:raggedtime}).
\end{itemize}

\section{Conclusion}
\label{sec:conclusion}

This paper has presented a highly parallel, fully GPU-resident, scalable
and efficient semi-Lagrangian drift-kinetic solver for tokamaks, on
unstructured poloidal planes of shaped cross section.  Its decomposition
divides the toroidal angle, the poloidal plane and the magnetic-moment
rows and leaves the parallel velocity whole, and its state never leaves
the GPU.  On Perlmutter's A100s a weak series that refines all five
phase-space dimensions and the time step together holds a weak-scaling
efficiency of $75\%$ on 16 GPUs and $57\%$ on 256, and four of the five
dominant kernels of the step run at a memory-bandwidth roof on a rank of
the 256-GPU point's shape.

The toroidal halo is deep, so the toroidal axis alone is capped at
$N_\varphi/s_w$ ranks, and its exchange sends each channel only its own
upwind depth; the results are bit-identical across process counts
and across nodes.  The $\mu$ rows, which no stage of the step couples,
multiply the rank count by the number of rows for one all-reduce of the
density per step, and the step strong-scales over them to 48 GPUs at
$61\%$ efficiency.  On the shaped plane a step costs within $5\%$ of one
on a circular plane of the same size.  At the strong series' size, the
companion paper's linear toroidal ITG demonstration on the circular cross
section, with every operator verified against an independently
computed reference~\cite{companion}, runs 1,200 steps to
$t = 120\,a/c_s$ on one A100 in under seven minutes, 332\,ms per step.
Every change was
held to a machine-checked gate with a stated exactness class
(Sec.~\ref{sec:method}).

Three costs that grow over the weak series remain: the stream apply,
which blocking over the plane should shorten; the field stage's
plane-sized work, which the row ranks could divide as they divide its
solves; and the waiting in the row collectives at 256 GPUs, which the
present design does not overlap.  Two extensions are planned: the weak
series on AMD MI250X GPUs, and its fourth point, the level-7 plane at 32
wedges on 4,096 GCDs.

\section*{Acknowledgments}

\textbf{AI-generated content.}  The design of
Secs.~\ref{sec:ragged}--\ref{sec:field}, the gate ladder and the
experiment harness were implemented, and the text of every section
drafted, with Anthropic's Claude~\cite{claude} (Fable~5 and~5.1, Opus~5
and~5.5), an AI code assistant working under \ifanon the
authors' \else the author's \fi direction; every section was edited by
\ifanon the authors\else the author\fi.  All results were reproduced from
the committed artifact, under the regression discipline of
Sec.~\ref{sec:method}, which we suggest --- a machine-checked exactness
class per change --- for AI-assisted performance engineering in general.
This paragraph is the disclosure of AI-generated text that the IPDPS
and IEEE policies require.
\ifanon

Further acknowledgments, among them funding and facilities, are withheld
for double-anonymous review.
\else

We thank Peng Wang for help with performance analysis.

This material is based upon work supported by the U.S. Department of
Energy, Office of Science, Office of Advanced Scientific Computing
Research, Scientific Discovery through Advanced Computing (SciDAC)
Program through the FASTMath Institute, under contract number
DE-AC02-05CH11231 at Lawrence Berkeley National Laboratory.  This
research used resources of the National Energy Research Scientific
Computing Center, a DOE Office of Science User Facility supported by
the Office of Science of the U.S. Department of Energy under Contract
No.\ DE-AC02-05CH11231 using NERSC award ASCR-ERCAP0039362.
\fi

\iflong
\appendices
\section{The gate ladder and what it found}
\label{app:gates}

This appendix is the long form of Sec.~\ref{sec:method}: the ladder
behind the four exactness classes, the three incidents the short version
does not report, and what the byte comparison of the restructuring does
and does not cover.

\subsection{The ladder}

The gate is a ladder of assertions with committed reference values, run
on every change: M0 to M17 in six groups, \emph{sharp} (the
machine-precision degenerations, the mesh and flux-surface labels, the
electron response, the row axis and the curved cells), \emph{orbit}
(bounce and transit frequencies), \emph{field} (Landau damping with the
mirror force, the polarization solve, the drift stage), \emph{drive}
(the local equilibrium, the E$\times$B source, the splitting order, the
envelope and helicity fits), \emph{itg} (the quantitative slab-ITG rate
and the dissipation floor) and \emph{mu} (the magnetic-moment
coordinate).  Every rung runs on the CPU build of the development machine and, as a
three-part batch on the 30-minute debug queue, on Perlmutter A100s
(parts of 21, 9 and 9 minutes at the current ladder), and a change is committed
only when both are green.

\subsection{Three further incidents}

\textbf{A term the degenerations cannot see.}  The E$\times$B source's
in-plane gradient was first taken at fixed perpendicular speed rather
than at fixed magnetic moment, an omission of a term tens of percent of
the drive at Cyclone Base Case gradients~\cite{dimits2000}.  It is
identically zero at uniform $B$,
so every sharp gate passed, and only an analysis of the operator exposed
it; the magnetic-moment row coordinate removes it by construction.  A
degeneration gate is blind to every term that vanishes in the
degeneration, and that is the reason the physics gates against a
toroidal reference exist.  \textbf{A lumped mass that is exactly zero.}
The averaging weight of the electron response was specified as the
lumped mass matrix, and on the $P_2$ space a vertex node's lumped
weight is exactly zero ($\int\varphi_i = 2A/12 - 2A/6 = 0$), so 161 of
the 577 nodes of the gate mesh carry no weight and a surface made only
of vertex nodes has none.  The first gate run stopped on the singular
surface block; the gate now prints the count of zero weights, and the
consistent mass matrix replaces the lumped one
(Sec.~\ref{sec:fsa}).  \textbf{A premise measured false before it was
built.}  The isoparametric boundary cells, an option on the circular
plane and not on the shaped one this paper runs, began from the
premise that the second-order error of the flux-surface average is the
sagitta of the straight cells.  Measured first, by sampling the field
exactly on the surfaces, the error is unchanged ($5.77\times10^{-2}$ to
$5.90\times10^{-2}$) and is the projector's own interpolation space; the
sagitta is first order and lives on the innermost ring, and the boundary
is the only place geometry costs an order.  The measurement replaced a
code plan for every cell with one for $1/K$ of them.

The 1X+3V predecessor supplied a fifth lesson, and it applies to the
shared stream machinery.  An exact-solution benchmark found that the
transport transfer had been built with a mis-paired overlap-matrix
convention: the operator transported cell means exactly, so every
conservation gate and damping fit passed, while the within-cell profile
carried $O(h)$ errors that reduced the transfer to first order.  A
ten-line test applying one sweep to $f(x) = x$ isolated the pairing,
and the fix is one line.  Conservation gates must be complemented by an
exact-solution order gate wherever one is available, which is why the
tokamak ladder asserts that in-space polynomials translate exactly
under the stream and the drift, and measures the interpolation order of
both under refinement.

\subsection{What a byte comparison covers}

The comparison behind the restructuring commits of
Sec.~\ref{sec:method} grew from 19 configurations to 29 over those
commits, and the reason generalizes.  A byte comparison covers only the
paths its configurations take.  The first stage we moved into an object
lost a default in the move --- its sub-cycle count became zero, so the
stage performed no work --- and all 19 configurations were
bit-identical, because that stage is active only under one option and
none of them set it.  Every object after it added a configuration that
sets its own options, including two that are bit-identical by
construction (a padded halo exchange and a device-direct one) and are
therefore invisible to any comparison that does not set them
deliberately.  The library the commits produced is 29 host units, 9
device units and 23 public headers, driven by a 239-line plain-C++
driver that never reaches the device compiler, with twelve objects that
carry create / set-from-options / set-up / view / destroy and own the
options that say what each one \emph{is}, while the run keeps those that
say how to compose it and what to check.  The finished library is held
to the whole ladder on the GPU nodes and to two frozen 100-step
histories of the production configuration, which it reproduces byte for
byte.

\section{On-chip velocity space on B200-class hardware}
\label{app:onchip}

Compute capability 10.0 provides 227\,KB of addressable shared memory per
SM, and thread-block clusters of 8 SMs portably (16 by opt-in), giving
1.8--3.6\,MB of distributed shared memory.  The five-species version of
this paper's Q3 velocity mesh (120 cells, 7{,}680 DOFs per species) is
$5\times7{,}680\times8\,\mathrm{B}=307$\,KB per spatial node, and a
$10^3$-cell Q3 mesh is $5\times64{,}000\times8\,\mathrm{B}=2.6$\,MB.  The
velocity sweep at one spatial node is independent of the others, so the
node is the unit a cluster holds: the first mesh fits a portable 8-SM
cluster and the second the 16-SM one.  A whole spatial cell of the
campaign's degree-2 elements (three nodes) is $0.92$\,MB on the first
mesh, inside an 8-SM cluster, and $7.7$\,MB on the second, which exceeds
both.  Spatial nodes then provide the grid-level parallelism and MPI ranks
the spatial decomposition.
\fi

\ifanon\else
\section*{Appendix: Artifact Description}

\begin{sloppypar}
\textbf{What the artifact is.}  \ifanon This description serves this paper
and its companion paper. \else This description serves two papers:
\emph{Parallel design of a GPU-resident semi-Lagrangian drift-kinetic solver
on unstructured extruded tokamak meshes} (the HPC paper) and \emph{KINTORA: a
semi-Lagrangian discontinuous Galerkin drift-kinetic, unstructured mesh
tokamak solver} (the methods paper).\fi{}  Their claims rest on one code base,
one gate ladder and one set of measurement files.  The code base is the
drift-kinetic solver, the \texttt{kin} library under \texttt{src/kin} and its
driver \texttt{ex11}, built on PETSc and Kokkos, with the 1X+3V
Vlasov--Poisson solver (\texttt{ex6}) and the 2X+2V constant-advection driver
(\texttt{ex8}) that the HPC paper's long version reports.  \ifanon The artifact
is a public repository, withheld here for double-anonymous review, \else The
artifact is the \texttt{kintora-hpc} and \texttt{kintora-methods} directories
of \url{https://gitlab.com/markadams4/petsc-claude-papers}, \fi which carry the
papers, the solver source frozen with the methods paper, the measurement data
and the scripts that reduce and plot it; each directory's \texttt{README}
records the commit.  Both papers' runs ran on earlier builds of the same tree:
the HPC paper's binaries are identified below by their checksums, and the
methods paper's data \texttt{README}s name the binary and source commit of
each run.  The reduced files the papers' numbers are
read from are in the artifact; the raw job outputs they were reduced from are
held in the run directories on the machine.

\textbf{Software.}  PETSc, \ifanon the authors' \else the author's \fi fork of
it, configured with the Kokkos and CUDA (\texttt{sm\_80}) backends; the fork is
needed for its device-side finite-element assembly and, to a lesser extent, its
device-side multigrid setup.  The GPU logs of both papers that carry PETSc's
revision line record the same commit of it, \texttt{80c04f7} of 2026-09-17, on
its public main branch, and the workstation gates build against the same
fork\ifanon{}
(the fork is withheld here for double-anonymous review)\else{}
(\texttt{petsc-claude} at \url{https://gitlab.com/markadams4/petsc-claude})\fi.
Kokkos as configured by PETSc; GPU-aware Cray MPICH; a GCC host compiler.
\texttt{src/kin/makefile} builds the library from its host and device units,
and the drivers' makefile links it.  Every option is listed under
\texttt{-help}.

\textbf{Hardware.}  Perlmutter (NERSC) GPU nodes: one AMD EPYC 7763 and four
NVIDIA A100s (40 and 80\,GB) per node on Slingshot-11, one MPI rank per GPU.
Perlmutter CPU nodes (two EPYC 7763) for the rank-invariance sweeps.
Development gates run on Apple M-series hardware with the Kokkos Serial
backend, which is also where the CPU half of every CPU-versus-GPU reproduction
is measured.

\textbf{Gates.}  The acceptance ladder is a committed script with committed
reference values, and it is what admits a change to the solver:
\texttt{scripts/check\_ex11\_vadvance.sh} on a workstation, and
\texttt{scripts/campaign/submit\_gate\_split.sh} on the machine, five
dependent jobs on the debug queue that take 82 minutes on one GPU node between
them, the longest 28.  Each physics gate asserts a measured rate against a
reference computed independently of the code, and the tools that compute those
references assert their own known degenerations before they are used.
\texttt{scripts/bytecheck\_ex11.sh} compares the physics output of two binary
sets byte for byte on 51 configurations; it is the arbiter of every commit
that moves code and changes no arithmetic.

\textbf{Experiment workflow.}  From the repository root,
\texttt{scripts/campaign/deploy.sh} copies the sources, meshes and scripts to
the machine by \texttt{rsync}, verifies every file by its checksum, and ships a
git bundle from which the gate ladder clones its historical rungs.  The series
binaries are built in a tree of their own and copied to the run directory, so a
build never races a gate.  Each timed run reports the PETSc
\texttt{-log\_view} events of its hot steps: a point of the strong and weak
series is 105 steps, of which the first five warm up in their own log stage,
except at the third weak-series point, which runs 30, and the ragged A/B runs
100 steps after one.  The runs use \texttt{-log\_view\_gpu\_time}.

\textbf{The HPC paper's measurements.}  The geometry of every timed run is
\texttt{-pol\_mesh\_file meshes/solovev\_L4.mesh -pol\_dm\_refine 0
-eq\_shape solovev}, with \texttt{solovev\_L5} and \texttt{solovev\_L6} at the
second and third weak-series points, and \texttt{-pol\_curved} off.  The
options the configuration turns on are \texttt{-nrow\_blocks} (the $\mu$ row
axis), \texttt{-field\_row\_split} (the per-plane solves divided among the row
ranks), \texttt{-field\_fsa} with \texttt{-fsa\_degree 2} (the electron
response), \texttt{-exb} (the E$\times$B term as a semi-Lagrangian stage),
\texttt{-phi\_pc\_type cholesky -phi\_pc\_factor\_mat\_solver\_type cudss}
(the direct per-plane solve), \texttt{-field\_batch\_stations} and
\texttt{-tor\_halo\_device\_mpi}, the last with
\texttt{MPICH\_GPU\_SUPPORT\_ENABLED=1} and \texttt{-{}-gpu-bind=none}.  The
one-sided worst-case exchange is the same binary with
\texttt{-tor\_halo\_worst\_case}, bit-identical in its physics to the ragged
one.  Each series has a comma-separated file under \texttt{data/} whose rows
carry the job they came from:
\begin{itemize}
\item the strong series over $\mu$ row blocks: \texttt{rowshaped\_submit.sh}
  over \texttt{ws\_x.sbatch}, jobs 58794292, 58794296, 58794298, 58794300 and
  58794302 with the solves divided, and 58804473, 58804475, 58804476 and
  58804477 with every row rank solving, a second allocation on a later binary;
  \texttt{row\_series.csv};
\item the accuracy-matched weak series: \texttt{ws16.sbatch} at
  \texttt{PT=A} and \texttt{PT=B}, jobs 58779096 and 58779097, with the
  field-stage entries from the device-timed reruns 58981406 and 58981408, and
  \texttt{ws16Cfix.sbatch}, job 58789831, at the third point;
  \texttt{weak\_series\_16x.csv};
\item what each optimization is worth: \texttt{ws16Cabl.sbatch}, job 58794289,
  with \texttt{ws16Bexb.sbatch}, job 58794291, for the E$\times$B row at the
  second point and \texttt{ws16Crep.sbatch}, job 58804472, for the
  divided-solve case; \texttt{ablation\_c.csv};
\item the roofline of the third point's rank: \texttt{rfC.sbatch}, job
  58877367, with the L2 and L1 columns from a second Nsight Compute run of the
  same binary, job 59180182; \texttt{roofline\_pointC.csv};
\item the ragged exchange in time: \texttt{raggedab.sbatch}, job 58867319, on
  the binary job 58867317 built; \texttt{ragged\_ab.csv};
\item the shaped plane's cost against the circular one at matched sizes:
  matched pairs of \texttt{ws16.sbatch}, jobs 58770209 to 58770212;
  \texttt{shaped\_vs\_circular\_stages.csv};
\item the long version's plane-count series (\texttt{ws\_x.sbatch} at
  \texttt{SERIES=plane}, job 58170363; \texttt{plane\_series.csv}), the A/B of
  the semi-Lagrangian E$\times$B stage against the Eulerian source (jobs
  58165464 and 58170789; \texttt{exb\_ab.csv}), the run-to-run repeats
  (\texttt{repeats.csv}) and the rank-invariance sweep
  (\texttt{cpu\_np\_sweep.sbatch}).
\end{itemize}
The binaries are \texttt{ex11\_ws19} (md5 \texttt{c3ec463a}), \texttt{ex11\_ws21}
(\texttt{ba45f3cf}), \texttt{ex11\_ws23} (\texttt{872086a1}) and
\texttt{ex11\_ws24} (\texttt{96dbebe5}), each built by its own job under
\texttt{scripts/campaign/}.  The GPU series take 27 node-hours together, and
the third weak-series point carries almost all of it: a run there holds 64
nodes, which makes the weak series $6.5$ node-hours and the ablation $20$,
against $0.44$ for the strong series in both columns.  Bit-identical
reproduction is expected for the rank sweep and for the row axis at one block;
the divided solves reproduce the every-rank ones to $2.9\times10^{-15}$, which
is the direct solver across devices; timing reproduces within the run-to-run
spread of a shared machine, measured at $3.0\%$ and $3.3\%$ of the mean on
repeated points.  Every point of the series is a single run in its own batch
allocation, so a difference below about $5\%$ between two single runs is not
an effect.

\textbf{The methods paper's figures and tables.}  Every figure has a committed
script that reads a committed data file and prints the numbers the text quotes:
\texttt{gen\_mms\_figure.py}, \texttt{gen\_bounce\_figure.py},
\texttt{gen\_orbit\_figure.py}, \texttt{gen\_landau\_figure.py},
\texttt{gen\_floor\_figure.py}, \texttt{gen\_coherence\_figure.py} and
\texttt{gen\_itg\_budget\_figure.py}, each named for its figure.  None is a
pixel-identical clone of the committed image, and two of them locate their
reference features, the envelope fit window and the recurrence time, from the
trace itself; each script's docstring says how.  The conservation and mass
tables are printed by \texttt{gen\_conservation\_tables.py} from the run logs,
and the table of local geodesic-acoustic roots by
\texttt{gen\_gam\_roots\_table.py}; the other appendix tables are typeset in
their \texttt{.tex} files, from the tools \texttt{scripts/ex11\_gam\_*.py},
\texttt{scripts/ex11\_zonal\_local\_ivp.py},
\texttt{scripts/ex11\_model\_residual.py} and
\texttt{scripts/ex11\_model\_observable.py}.  The toroidal ITG rates are the
legs of \texttt{scripts/campaign/cbc6.sbatch}, one A100 each, with a
\texttt{README} in \texttt{data/} that maps every leg to its job; the
demonstration is job 59271321, 1,200 steps in 532\,s with its halo staged
through the host; its repeat with the device-direct halo, job 59483229,
reproduces it to the printed digits in 413\,s, 332\,ms per step, and is the
timing the papers quote.

\textbf{The zonal runs.}  The zonal-flow measurements of the methods paper run
on GPU nodes through \texttt{scripts/campaign/rhtor.sbatch} and the runners
beside it that sweep the toroidal wedge count (\texttt{rhtorN.sbatch}, and
\texttt{rhcpuN.sbatch} on CPU nodes), run the default step as a control
(\texttt{rhtor\_def.sbatch}) and pair a circular control with the shaped plane
(\texttt{rh\_shaped8.sbatch}, which fixes both); the others set the seed
amplitude and the electron closure through their environment.
Every one of those runs is identified by its wedge count, its seed amplitude,
its closure and its step variant, and no number from them can be read without
all four: the closure moves the geodesic-acoustic damping by a factor of $1.5$
to $1.7$, by fit window, two wedges read the residual 38\% to 44\% below eight,
and the seed amplitude moves the damping by $9\%$ and the unsubtracted residual
by a factor of four.  The torus's flux-surface average in the electron closure
is the code's default; a driver built before that default must be given the
option explicitly.

\textbf{Known gaps.}  The right panel of the slab-ITG budget
figure, a four-configuration control set, has no data file; its left panel is
regenerated in full.  Two figures built earlier in the work are committed but
referenced by neither paper, and one of them cannot be regenerated because the
options of its run were not recorded.  The batch scripts carry the project
account and absolute paths of the machine they ran on, and must be edited
before they run elsewhere.
\end{sloppypar}

\fi
\end{document}